\documentclass[trackchanges]{aastex701}

\usepackage{stackengine}
\setstackgap{S}{1pt}
\setstackgap{L}{.7\baselineskip}

\begin{document}

\title{Observations of a Twin Pair of Atypical Solar Flares and a Magnetic-reconnection Scenario}

\author[orcid=0000-0002-0833-9485,sname='Agarwal']{Satyam Agarwal}
\affiliation{Center for Space Plasma and Aeronomic Research, The University of Alabama in Huntsville, 35899, Huntsville, AL, USA}
\email[show]{satyamagarwal33@gmail.com}  

\author[orcid=0000-0003-1281-897X]{Alphonse C. Sterling} 
\affiliation{NASA Marshall Space Flight Center, Huntsville, AL 35812, USA}
\email{alphonse.sterling@nasa.gov}

\author[orcid=0000-0002-9370-2591]{Shibu K. Mathew}
\affiliation{Udaipur Solar Observatory, Physical Research Laboratory, Udaipur, Rajasthan, 313001, India}
\email{shibu@prl.res.in}

\author[orcid=0000-0002-5691-6152]{Ronald L. Moore}
\affiliation{Center for Space Plasma and Aeronomic Research, The University of Alabama in Huntsville, 35899, Huntsville, AL, USA}
\affiliation{NASA Marshall Space Flight Center, Huntsville, AL 35812, USA}
\email{ronald.l.moore@nasa.gov}

\author[orcid=0000-0002-7570-2301]{Qiang Hu}
\affiliation{Center for Space Plasma and Aeronomic Research, The University of Alabama in Huntsville, 35899, Huntsville, AL, USA}
\affiliation{Department of Space Science, The University of Alabama in Huntsville, Huntsville, AL 35899, USA}
\email{qh0001@uah.edu}

\author[orcid=0000-0003-4522-5070]{Ramit Bhattacharyya}
\affiliation{Udaipur Solar Observatory, Physical Research Laboratory, Udaipur, Rajasthan, 313001, India}
\email{ramit@prl.res.in}



\begin{abstract}
We present observations and a magnetic-reconnection scenario of a twin pair of ``atypical flares" that occurred on 2022 April 22 in a quadrupolar magnetic configuration formed by two active regions. The spatio-temporal evolution of the two flares is examined using images from the Atmospheric Imaging Assembly (AIA) onboard the Solar Dynamics Observatory (SDO), and from the ground-based Multi-Application Solar Telescope (MAST) in Udaipur, India. Characteristic of atypical flares and indicative of slipping reconnection, the ribbons of each of our atypical flares (1) do not spread apart, and (2) grow longer by sequential brightening of new flare kernels.
The two atypical flares are homologous and plausibly have homologous triggers. There are four additional pairs of flare ribbons, each from a different flaring event that releases much less energy than the atypical flares. Two of these four pairs are each made by a precursor, each possibly triggering one of the two atypical flares. The remaining two pairs accompany a filament's activation, occurring twice within the span of the two atypical flares. Using a nonlinear force-free field (NLFFF) extrapolation model, we approximated the coronal magnetic field and found two quasi-separatrix layers (QSLs) that are nearly rooted in the flare ribbons of the atypical flares. The observations and the extrapolated field together suggest a scenario in which the nearly simultaneous occurrence of many reconnections between magnetic field lines crossing each other at small angles (slipping reconnection) \textit{within} each of the two QSLs makes the observed pair of atypical flares.
\end{abstract}




\section{Introduction} 
Solar flares are explosive events on the Sun made by a sudden release of magnetic energy. These events occur when magnetic reconnection occurs in the solar atmosphere, converting stored magnetic energy into heat, kinetic energy of the magnetofluid, and acceleration of charged particles \citep{2011S}. Solar flares are broadly of two types: eruptive and confined. Solar flares that are accompanied by coronal mass ejections (CMEs) are called ``eruptive flares," whereas those that are not accompanied by CMEs are called ``confined flares" \citep{Svestka1992}. This latter category, confined flares, may be further subcategorized into two varieties. The first of these are flares that start off like typical CME-producing eruptive flares, but where the eruption is not strong enough to escape the confining coronal field (e.g., \citealp{2001ApJ...552..833M,2023MNRAS.525.5857J}). Because this type of eruption can show a filament that first starts to erupt and then suddenly stops, this type of confined flare is sometimes called a ``failed eruption" \citep{2003ApJ...595L.135J}. In recent years, however, it has become apparent that such failed eruptions are not the only type of flares lacking an accompanying CME. This second, newly recognized subcategory of confined flares are called ``atypical flares," in which magnetic energy is released by reconnection but where there is no noticeable erupting structure and no CME. To date, atypical flares have been little studied, and they are the topic of this paper.


In the well-known two-ribbon solar flares, two parallel flare ribbons form on each side of the polarity inversion line (PIL) in the photospheric magnetic flux. The ribbons are connected by a growing magnetic arcade and move away from each other during the flare. These types of flares are well-explained by the two-dimensional (2D) standard flare model, also known as the CSHKP \citep{carmichael1964,sturrock1966,hirayama1974,kopp1976} model. In that model, an erupting filament (or prominence) results in reconnection below it in a current sheet. Charged particles are accelerated by the reconnection and move downward along the legs of the magnetic arcade made by the reconnection, heating the flare ribbons and post-flare loops. The site of reconnection keeps moving upward under the erupting structure, and the two ribbons move apart. For further details, see \citet{2011S} and \citet{2022SoPh..297...59K}.


From the viewpoint of modeling, instability of a magnetic flux rope (MFR) existing prior to the eruption, and reconnection between sheared field lines leading to the formation of a MFR early in the eruption, are considered to be the primary mechanisms initiating eruptive flares and failed eruptions \citep{2017ScChD..60.1383C,2024ScChD..67.3765J}. Two well-known examples of these mechanisms are the torus instability \citep{2006PhRvL..96y5002K} and tether-cutting reconnection \citep{2001ApJ...552..833M} models.
Furthermore, a standard flare model in 3D has been proposed by \citet{2012A&A...543A.110A,2013A&A...549A..66A} and \citet{2013A&A...555A..77J}, incorporating the physics of reconnection of a pair of quasi-separatrix layers (QSLs; \citealp{2002TH}) that envelop an erupting MFR. The model consists of a torus-unstable MFR, whose eruption results in magnetic reconnection below it at a hyperbolic flux tube (HFT; \citealp{2002TH}) made by the intersection of the two QSLs. The photospheric footprints of the two QSLs are J-shaped on either side of the PIL. The two hooked parts of the two J-shapes encircle the MFR feet, and the two straight parts become the two flare ribbons. The formation of post-flare loops and growth of the erupting MFR occurs by a continuous series of 3D magnetic reconnections in the current sheet under the erupting MFR. In some studies, this model has been used to explain flare observations. For example, \citet{Janvier_2014} used it to account for the observed J-shaped flare ribbons in their study. Also, \citet{Zhao_2016} used the model to explain the formation of an S-shaped sigmoid due to reconnection between double J-shaped quasi-separatrix layers.

All flares that involve an erupting magnetic flux rope can be called ``standard flares." If the flux rope escapes into the heliosphere, then the eruption is ejective.  If it does not escape, then it is a confined eruption of the ``failed eruption" subcategory. Interestingly, recent studies have reported solar flares that cannot be explained by any of the the existing solar flare models. These are the ``atypical flares". In these flares, the flare ribbons brighten suddenly in place and do not spread apart, in contrast to standard two-ribbon flares.

The first study of atypical flares by \citet{2015A&A...574A..37D} found a confined GOES C3.3 that they argued could not be explained by a standard flare model. They reported the development of two flare ribbons parallel to a pair of filaments. Post-flare loops arched over the filaments, but those filaments apparently did not play any role in the flare's formation or evolution. The authors argued that these features are inconsistent with the standard model, and they therefore referred to it as an atypical flare. They found the magnetic configuration to consist of a complex system of many QSLs in which multiple and sequential reconnections produce the flare.


In another study, \citet{2019ApJ...871..165J} analyzed a long-duration confined C-class flare and found similar results as \citet{2015A&A...574A..37D} with respect to the formation of the post-flare loops, the role of the QSLs, and stability of the filaments. They also proposed a picture for the flare mechanism wherein quasi-parallel crossed magnetic field lines across a curved PIL reconnect with each other. 

In a statistical study, \citet{2019ApJ...881..151L} investigated a total of 18 confined solar flares and found them to be of two types, ``Type I" and ``Type II." The ``Type I" flares are characterized by slipping reconnections in multiple QSLs, a stable filament, strongly sheared post-flare loops arching over the non-eruptive filament, and preservation of the overall magnetic configuration. They recognized the ``Type I" flares to be essentially the same as the atypical flare of \citet{2019ApJ...871..165J}. For``Type II" flares, they found weakly sheared flare loops in a filament eruption that is confined by strong strapping due to the overarching field, and they concluded that ``Type II" flares fit the standard model for confined flares. In a more recent study, \citet{2022ApJ...933..191D} investigated a total of 152 confined flares and found that nearly $40\%$ of the flares belonged to the ``Type I" category, while the rest belonged to the ``Type II" category. 


While there has been much work expended in understanding the mechanism of standard flares, there so far have been few detailed examples of atypical flares, and there still is much uncertainty about the mechanism that powers them.  With this in mind, we investigate a twin pair of atypical solar flares in this study. The two atypical flares are found to occur in nearly the same magnetic field arches, each making nearly the same two flare ribbons that do not spread apart. The observations and analysis show a filament that does not play any role in the dynamics of the two atypical flares. That filament becomes activated during each of the two atypical flares, but does not display the dynamic activity that normally stands out in observations of typical standard eruptions. Furthermore, the flare loops arch over the filament, in contrast to failed filament eruptions where the flare arcade forms below the rising filament before its outward motion is stopped by the confining coronal field. From the observations and a nonlinear force-free field model of the coronal magnetic field, we propose that each of the two flares results from slipping reconnection in magnetic loops rooted in the flare ribbons.

\section{Observations} \label{region-flare}

\begin{figure*}[h]
\epsscale{0.9}
\plotone{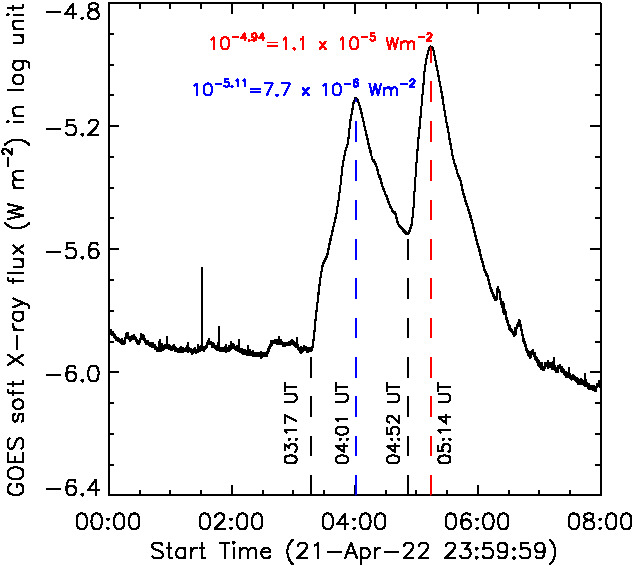}
\caption{Temporal evolution of X-ray irradiance measured by the X-Ray sensor of the GOES-17 satellite in the $0.1-0.8$ nm bandpass at 1-second cadence for a duration of 8 hours. The dashed black lines mark the start times of the GOES C7.7 and GOES M1.1 flares, while the dashed blue and red lines mark their respective peak times. The $y$-axis uses log scaling. The texts in blue and red are the peak X-ray fluxs in units of $\rm{W\,m^{-2}}$ for the first and second atypical flares, respectively.
\label{goes}}
\end{figure*}

We investigate a twin pair of homologous atypical solar flares observed on 2022 April 22 in the magnetic configuration formed by adjacent bipolar NOAA active regions (ARs) 12993 and 12994. The ARs each had a $\beta\gamma$ configuration on April 22. The pair had a complex arrangement of magnetic flux with intermixed polarities. The twin flares were of different strengths, classed as GOES C7.7 and GOES M1.1 flares by their measured peak soft X-ray (SXR) emission. \autoref{goes} shows the temporal evolution of the SXR irradiance, measured by the X-Ray Sensor (XRS) of the GOES-17\footnote{\url{https://www.ncei.noaa.gov/products/goes-r-extreme-ultraviolet-xray-irradiance}} satellite in the wavelength range of $0.1-0.8$ nm at a cadence of one second.

The plot shows a sudden rise in the SXR flux at 03:17 UT, marking the beginning of the impulsive phase of the first flare. The SXR flux peaks at $\rm{7.7\times10^{-6}\,W\,m^{-2}}$ at 04:01 UT, making the flare a GOES C7.7 flare. Then, the emission decreases until 04:52 UT, when it starts increasing due to the impulsive phase of the second flare. The flux reaches its maximum of $\rm{1.1\times10^{-5}\,W\,m^{-2}}$ at 05:14 UT, making the flare a GOES M1.1 flare. Thereafter, the flux decreases and eventually returns to its pre-flare level. Since the second flare occurs during the decay phase of the first flare and has nearly the same flare ribbons, we refer to this pair of flares as a twin pair. Because the two flares have nearly the same flare ribbons, we call these flares homologous. To investigate further, we used photospheric magnetograms and multi-wavelength images of the flares, as described in the following sections.

\subsection{Photospheric Magnetic Field\label{subsec:pmf}}

\begin{figure*}[h]
\epsscale{1.0}
\plotone{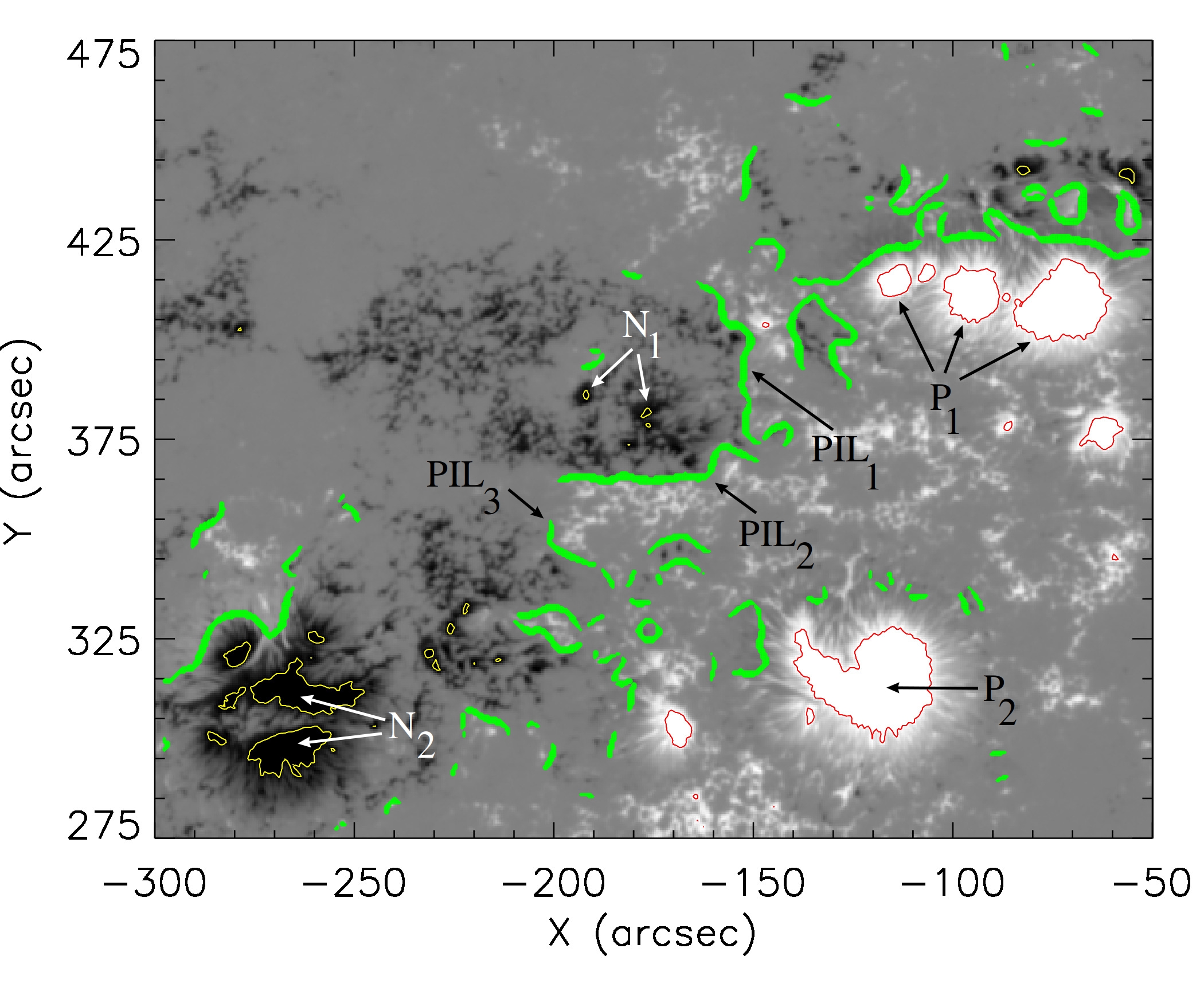}
\caption{A cutout of a full-disk line-of-sight magnetogram showing the two active regions, scaled at $\pm 1000$ G, with red and yellow contours at the level of  $+1000\,\mathrm{G}$ and $-1000\,\mathrm{G}$, respectively. The positive ($\rm{P_{1}}$, $\rm{P_{2}}$) and negative ($\rm{N_{1}}$, $\rm{N_{2}}$) polarities at 03:00 UT on 2022 April 22 are pointed to by the arrows. The green paths are polarity inversion lines, three of which are pointed to and labeled as $\mathrm{PIL_1}$, $\mathrm{PIL_2}$, and $\mathrm{PIL_3}$. North is upward and west is to the right in this image, and in all other solar images in this paper.
\label{mag}}
\end{figure*}

We used line-of-sight magnetograms from the Helioseismic and Magnetic Imager (HMI; \citealp{2012Schou}) onboard the Solar Dynamics Observatory (SDO; \citealt{2012P}) to analyze the magnetic configuration of the adjacent active regions. \autoref{mag} shows a cut-out from the full disk magnetogram at 03:00 UT on 2022 April 22, obtained from hmi.M\_720s data series of SDO/HMI.
In the north of the figure, closely-spaced regions (sunspots) of strong positive polarity (pointed to by arrows and collectively labeled $\rm{P_{1}}$) and two sunspot pores of strong negative polarity (pointed to by arrows and collectively labeled $\rm{N_{1}}$) are visible. Together, $\rm{P_1}$ and $\rm{N_1}$ constitute active region AR 12993, whose positive polarity is still mostly in full-fledged sunspots, while the negative polarity has decayed significantly and no longer has sunspots. In the south of the figure, strong concentrations of positive and negative polarities (labeled $\rm{P_{2}}$ and $\rm{N_{2}}$, respectively) constitute active region AR 12994, where both polarities are mostly in full-fledged sunspots. The two active regions together form a quadrupolar magnetic configuration, and since this configuration has only the three clusters of full-fledged sunspots, we describe it as asymmetric. This asymmetry in the distribution of magnetic flux is reflected in the connection of magnetic field lines between the magnetic polarities, as will be shown later. The magnetogram also reveals extended regions of unipolar plage, such as between $\rm{P_{1}}$  and $\rm{P_{2}}$, surrounding $\rm{N_{1}}$, and adjacent to $\rm{N_{2}}$ toward its northwest. In consideration of this, we further characterize the quadrupolar configuration as fragmented, in addition to being asymmetric. The figure shows the PILs in green, with three of them labeled. The three labeled PILs collectively make an overall inverse S-shaped PIL between the positive and negative flux north of the southern active region (AR 12994).

\begin{figure*}[h]
\epsscale{1.0}
\plotone{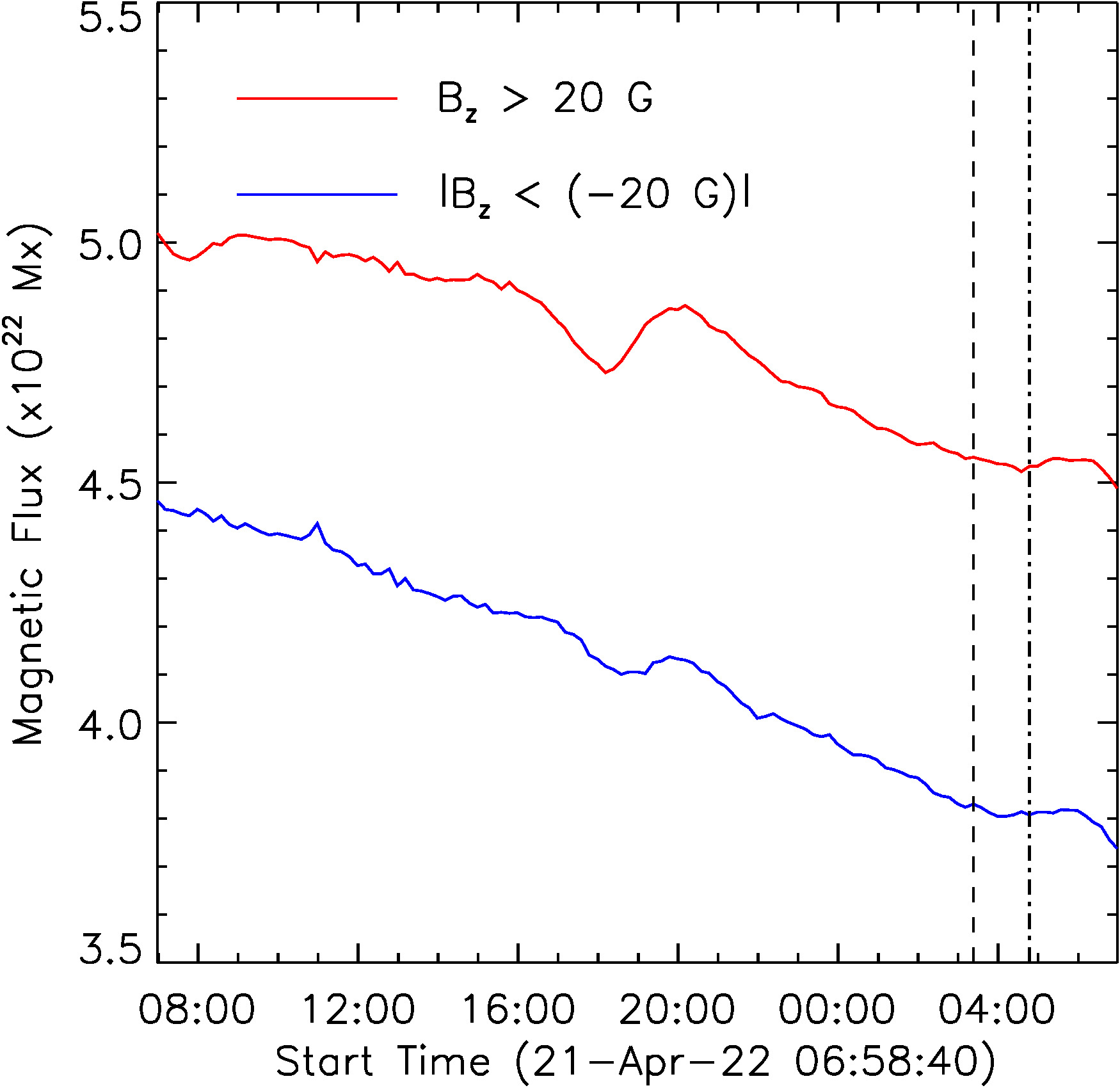}
\caption{Evolution of the positive (red) and negative (blue) magnetic flux in the field of view (FOV) of \autoref{mag} for a duration of 24 hours, starting from 2022 April 21, 07:00 UT. The dashed and dot-dashed lines mark the starting times of the two solar flares.
\label{magflux}}
\end{figure*}

We also investigated the evolution of the photospheric magnetic flux, which can play an important role in triggering solar flares that involve MFR dynamics \citep{2000ApJ...545..524C,2018ApJ...864...68S,2024ScChD..67.3765J}, but its role in the triggering of atypical flares remains largely unexplored. In this study, we used the hmi.sharp\_cea\_720s series, which provides the vector magnetic field and uses cylindrical equal area projection, which basically means that every pixel is equal in solar surface area. The rationale for using the vector magnetic field is that esimating the flux from the line-of-sight field component does not account for the transverse magnetic field and is therefore less accurate \citep{2017SoPh..292...36L}. Therefore, the hmi.sharp\_cea\_720s series is ideal for magnetic flux measurement. The magnetic field components of this data series are given as $\mathrm{B}_{r}$ ($r$; radial), $\mathrm{B}_{p}$ ($p$; poloidal), and $\mathrm{B}_{t}$ ($t$; toroidal) in heliocentric spherical coordinates, and are related to the heliographic components as $\mathrm{B}_{z}$=$\mathrm{B}_{r}$, $\mathrm{B}_{x}$=$\mathrm{B}_{p}$, and $\mathrm{B}_{y}$=$\mathrm{-B}_{t}$. We used the $\mathrm{B}_{z}$ component to follow the evolution of positive and negative magnetic flux in the area shown in \autoref{mag}. In \autoref{magflux}, we plot the evolution of flux in that heliographic area for a duration of 24 hours, starting from 2022 April 21, 07:00 UT. The plot shows that for that area, there is a net decrease, amounting to $5.31\,(7.25)\times 10^{21}$ Mx for the positive (negative) flux. Furthermore, the change in the magnetic flux near the boundary of that field of view is found to be two orders of magnitude smaller than for the whole area, showing that flux cancellation dominates over both flux emergence and transport of flux to outside that area by photospheric flows. Importantly, although this implies that flux cancellation dominates the evolution of the photospheric magnetic flux in the active-region area that hosted the observed atypical flares, we could not discern the role of that cancellation, if any, in the triggering of the atypical flares.
\subsection{Multi-wavelength Imaging Observations}
\label{observation}
\begin{figure*}[h]
\begin{interactive}{animation}{Figure4.mp4}
\epsscale{0.9}
\plotone{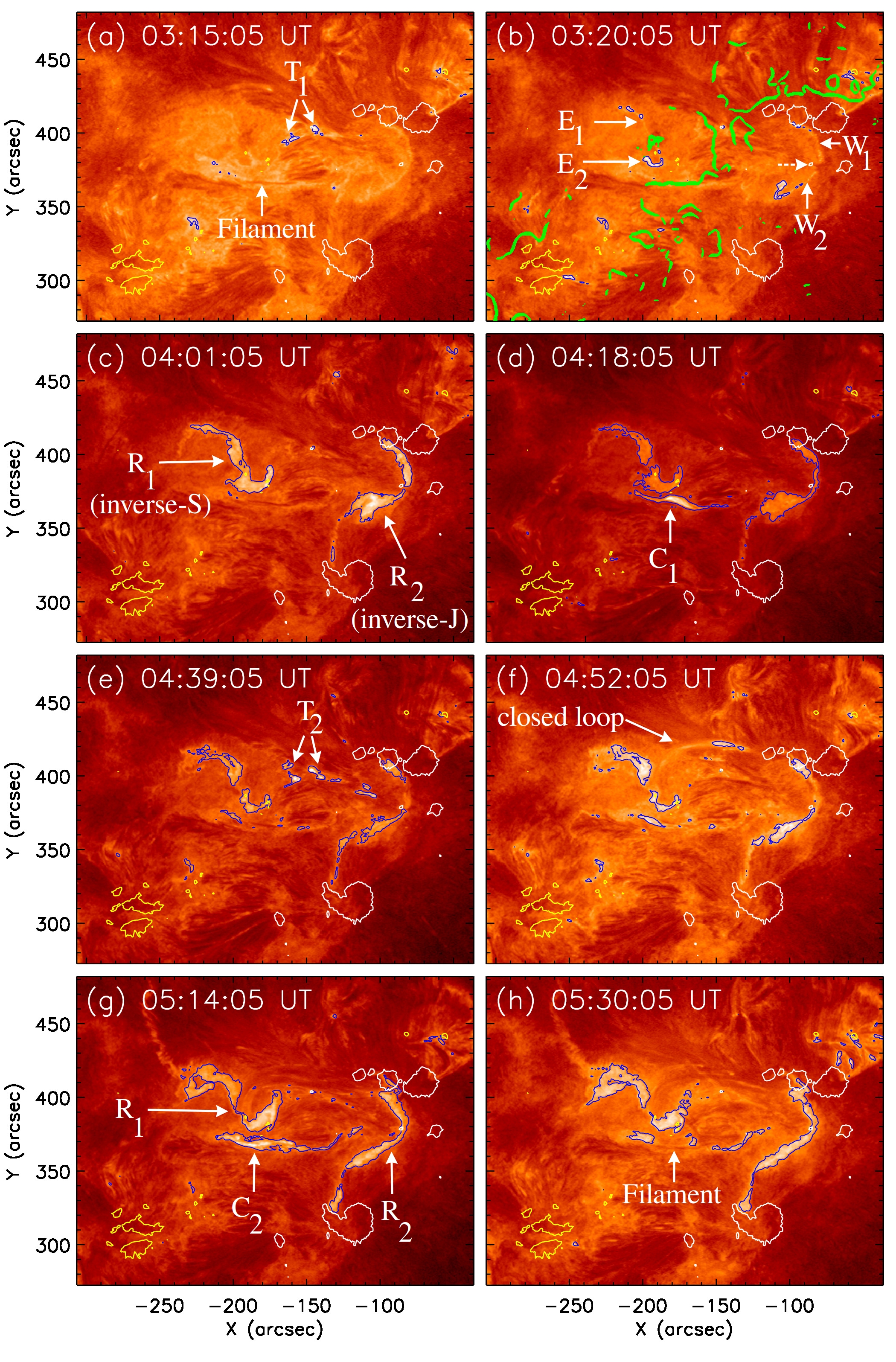}
\end{interactive}
\caption{Sequence of SDO/AIA 304\,\AA\,images having the FOV of \autoref{mag}: $\mathrm{R_{1}}$ and $\mathrm{R_{2}}$ are the two ribbons of the twin flares; $\mathrm{C_1}$ and $\mathrm{C_2}$ are brightenings along $\mathrm{PIL_2}$ at 04:18 UT and 05:14 UT; and $\mathrm{T_1}$ and $\mathrm{T_2}$ are brightenings bracketing a northern segment of $\mathrm{PIL_1}$ at 03:15 UT and 04:39 UT. $\mathrm{E_1}$, $\mathrm{E_2}$ and $\mathrm{W_1}$, $\mathrm{W_2}$ are two early segments of $\mathrm{R_1}$ and $\mathrm{R_2}$, respectively. The green contours and the dashed white arrow in panel (b) trace the PILs and point to the clump of positive-polarity flux separating $\mathrm{W_1}$ and $\mathrm{W_2}$. The filament is pointed to in panels (a) and (h). A closed magnetic loop rooted in $\mathrm{R_1}$ and $\mathrm{R_2}$ is pointed to in panel (f). The white (yellow) contours are for $\mathrm{B}_z$=$\mathrm{+1000\,(-1000)\,G}$ at 03:00 UT, while the blue contours trace brightness intensity level 100 DN.\\ \textbf{Animation:} An animation showing the SDO/AIA 304\,\AA\,images from 02:00 UT to 06:00 UT at a cadence of 12 seconds is available in the HTML version of this article.
\label{304}}
\end{figure*}

To examine the spatio-temporal evolution of the two homologous atypical flares, we used images in the 304\,\AA\, and 131\,\AA\, wavelength channels of the Atmospheric Imaging Assembly (AIA; \citealp{2012L}) onboard SDO, and in the Ca II 8542\,\AA\, line of the ground-based Multi-Application Solar Telescope (MAST; \citealt{2017SoPh..292..106M}) on an island observatory in a lake in Udaipur, India. The observations reveal multiple pairs of flare ribbons. Importantly, the pairs of flare ribbons of the two atypical flares are nearly the same, and in \autoref{304}, we label the flare ribbons of these pairs $\mathrm{R_1}$ and $\mathrm{R_2}$ for both atypical flares. In addition to $\mathrm{R_1}$ and $\mathrm{R_2}$, we identify four different pairs of flare ribbons (labeled $\mathrm{T_1}$, $\mathrm{T_2}$, $\mathrm{C_1}$, and $\mathrm{C_2}$), each from a different flaring event that releases much less energy compared to the atypical flares. We choose these labels because, as described later, the events that made $\mathrm{T_1}$ and $\mathrm{T_2}$ possibly ``trigger" the two atypical flares, and because $\mathrm{C_1}$ and $\mathrm{C_2}$ each accompany the activation of a filament that remains ``confined" during the two atypical flares.
The details of these four pairs of ribbons and their relevance to the evolution of the two atypical flares will be discussed shortly. The observations also reveal a filament along $\mathrm{PIL_2}$ and hot coronal loops whose footpoints at one end are rooted in ribbon $\mathrm{R_1}$ and the other end in ribbon $\mathrm{R_2}$. In the following, we discuss the spatio-temporal evolution of the above-mentioned features in detail.

\autoref{304} shows images of the two homologous atypical flares in the 304\,\AA\,channel of SDO/AIA at sequential times. \autoref{304}(a) points out the filament and shows a pair of flare ribbons  (labeled $\mathrm{T_1}$) that shortly precedes the onset of the impulsive phase of the first atypical flare. The ribbons in $\mathrm{T_1}$ remain visible for a few minutes only and are soon followed by sudden, non-uniform flare brightenings in the negative-polarity plage around $\mathrm{N_1}$ and in the positive-polarity plage between $\mathrm{P_{1}}$ and $\mathrm{P_{2}}$, as seen in \autoref{304}(b). These brightenings are early in the ribbons of the first atypical flare, which are labeled $\mathrm{R_1}$ and $\mathrm{R_2}$ in \autoref{304}(c). Notably, in the onset of the first atypical flare, $\mathrm{R_1}$ and $\mathrm{R_2}$ are disjoint, each consisting of two distinct bright segments: $\mathrm{E_{1}}$ and $\mathrm{E_{2}}$ for $\mathrm{R_1}$, and $\mathrm{W_{1}}$ and $\mathrm{W_{2}}$ for $\mathrm{R_2}$. The segments $\mathrm{E_1}$ and $\mathrm{E_2}$ are separated by a PIL, shown in green, while $\mathrm{W_1}$ and $\mathrm{W_2}$ are separated by a clump of positive polarity flux, pointed to by a dashed white arrow in \autoref{304}(b). These segments grow in place and lengthen such that $\mathrm{R_{1}}$ and $\mathrm{R_{2}}$ exhibit an inverse-S and an inverse-J shape at the peak time of the first atypical flare, as shown in \autoref{304}(c). In the decay phase of the first atypical flare, a separate flare brightening (labeled $\mathrm{C_1}$) that remains discernible for nearly 5 minutes is observed across and along $\mathrm{PIL_2}$ at 04:18 UT, as shown in \autoref{304}(d). Notably, although not immediately evident in the images, $\mathrm{C_1}$ consists of a pair of flare ribbons bracketing $\mathrm{PIL_2}$. The difficulty in discerning these two ribbons is presumably due to the compactness of the eruption that made $\mathrm{C_1}$ and due to the lack of substantial movement of these ribbons. During the evolution of the first atypical flare (flare ribbons $\mathrm{R_{1}}$ and $\mathrm{R_{2}}$) there occur two additional transient pairs ($\mathrm{T_{1}}$ and $\mathrm{C_{1}}$) of flare ribbons resulting from two separate flaring events distinct from the first atypical flare.


Following the first atypical flare, another pair of ribbons (labeled $\mathrm{T_2}$) appears at 04:39 UT at the site of $\mathrm{T_1}$. The flare ribbons in $\mathrm{T_2}$ remain visible for about 3 minutes and are shown in \autoref{304}(e). Notably, the timings of $\mathrm{T_1}$ (03:15 UT) and $\mathrm{T_2}$ (04:39 UT) relative to the start times of the two atypical flares, along with their spatial position with respect to $\mathrm{R_1}$ and $\mathrm{R_2}$, suggest that the energy-release events that made the pairs $\mathrm{T_1}$ and $\mathrm{T_2}$ triggered the first and second atypical flares of the twin pair, respectively. Following \citet{2017NatAs...1E..85W}, we call these events ``precursors," each of which possibly triggered one of the observed atypical flares. While the triggering mechanisms of the two atypical flares remain uncertain, and while $\mathrm{T_1}$ and $\mathrm{T_2}$ might have or might not have triggered the two atypical flares, $\mathrm{T_1}$ and $\mathrm{T_2}$ are reasonable possibilities, particularly because of their timing and location relative to those of the atypical flares. Moreover, we could not identify any other possible triggers for the observed atypical flares. Following $\mathrm{T_2}$, the ribbons of the second atypical flare start to brighten, as seen in \autoref{304}(f). Additionally, a cooling magnetic loop (presumably a post-flare loop of the first flare) having footpoints in $\mathrm{R_{2}}$ and segment $\mathrm{E_2}$ of $\mathrm{R_{1}}$ becomes visible, as seen in that same panel. The image of this loop suggests that it arches higher than the height of the filament---a characteristic of atypical flares. Importantly, the ribbons of the second atypical flare are observed to be cospatial with the flare ribbons of the first atypical flare, indicating that the two flares occurred in nearly the same magnetic field arches. We label the ribbons of the second flare also $\mathrm{R_1}$ and $\mathrm{R_2}$ in \autoref{304}(g). This panel also shows a flare brightening (labeled $\mathrm{C_2}$) that is cospatial with $\mathrm{C_1}$. This brightening forms due to another energy-release event and appears about 15 minutes after the start time of the second atypical flare. Similar to $\mathrm{C_1}$, this brightening also consists of a very close pair of flare ribbons. The coronal emission from the second atypical flare then fades as the ribbons $\mathrm{R_1}$ and $\mathrm{R_2}$ become faint. Importantly, the filament remains identifiable at least until 05:30 UT (\autoref{304}(h)), and the filament does not exhibit any notable rise, remaining confined throughout the duration of the flaring dynamics---a characteristic of atypical flares. Additionally, since both $\mathrm{C_1}$ and $\mathrm{C_2}$ are along the filament's PIL, the confinement of the filament suggests that its activation was triggered by either flux cancellation at $\mathrm{PIL_2}$ or by the ongoing atypical flares, resulting in $\mathrm{C_1}$ and $\mathrm{C_2}$. Similar to the first atypical flare, the duration of the second atypical flare spans flare ribbons $\mathrm{R_{1}}$ and $\mathrm{R_{2}}$, along with two additional and transient pairs ($\mathrm{T_{2}}$ and $\mathrm{C_{2}}$) of flare ribbons resulting from two different events of energy release.

\begin{figure*}[h]
\begin{interactive}{animation}{Figure5.mp4}
\epsscale{1.0}
\plotone{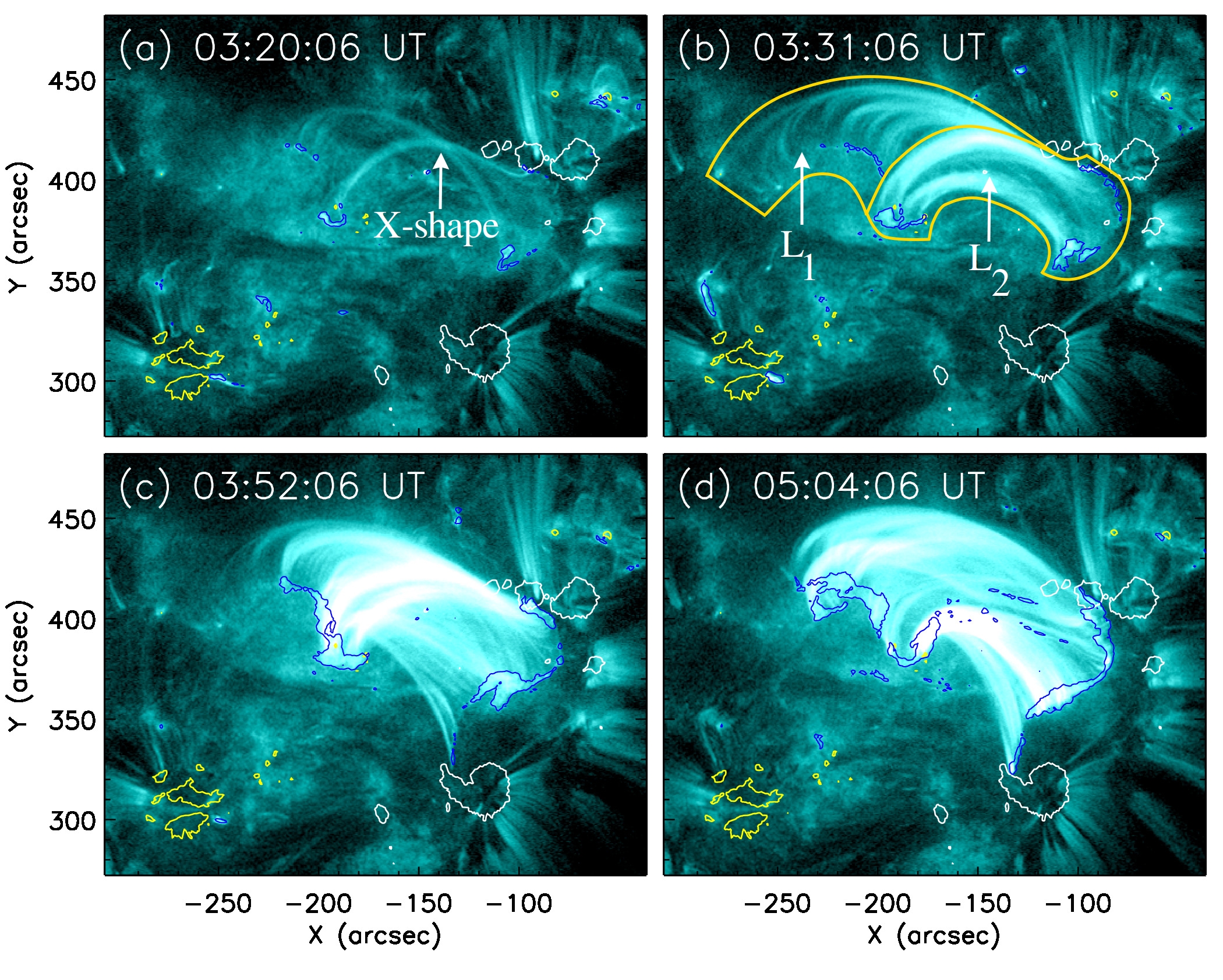}
\end{interactive}
\caption{Sequence of SDO/AIA 131\,\AA\,images of the twin flares. Panel (a) shows the apparent X-shape, and panel (b) shows the two loop systems $\mathrm{L_1}$ and $\mathrm{L_2}$, outlined by the gold contours. Panels (c) and (d) show each flare near its peak time. The white (yellow) contours trace $\mathrm{B}_z$=$\mathrm{+1000\,(-1000)\,G}$ at 03:00 UT, while the blue contours outline the SDO/AIA 304\,\AA\,flare ribbons at emission intensity level 100 DN.\\ \textbf{Animation:} An animation showing the SDO/AIA 131\,\AA\,images from 02:00 UT to 06:00 UT at a cadence of 1 minute is available in the HTML version of this article.
\label{131}}
\end{figure*}

We also studied the images from the 131\,\AA\,channel (\autoref{131}) of SDO/AIA to investigate the evolution of the hot flare arcades in the solar corona. In these images, the GOES C7.7 flare starts with the appearance of two faint loops forming an X-shape, as shown in \autoref{131}(a). In this apparent X-shape, one loop of the ``X" is rooted in the vicinity of $\mathrm{P_1}$ and $\mathrm{N_1}$, while the other loop has its footpoints in the plage northeast of $\mathrm{N_1}$ and in the plage between $\mathrm{P_1}$ and $\mathrm{P_2}$. However, we cannot tell whether these two loops are well apart along the line of sight through their crossing point in the image, or instead are in contact there and so might be reconnecting. Subsequently, additional brighter loops appear, as seen in \autoref{131}(b). The morphological evolution of these loops suggests that they can be categorized into two distinct bundles or fans of loops. These fans, labeled $\mathrm{L_1}$ and $\mathrm{L_2}$, are outlined in \autoref{131}(b) by gold contours. The positive feet of the loops in $\mathrm{L_1}$ are compactly clustered near $\mathrm{P_1}$, while at the opposite (negative) end, they fan out into the plage near and northeast of $\mathrm{N_1}$. The brightening of loops in $\mathrm{L_1}$ is accompanied by sequential brightening of flare kernels eastward along segment $\mathrm{E_1}$ of ribbon $\mathrm{R_1}$, which indicates slipping reconnection---a characteristic of atypical flares. The negative ends of the loops in $\mathrm{L_2}$ are in segment $\mathrm{E_2}$ of ribbon $\mathrm{R_1}$ and the positive ends are in segments $\mathrm{W_1}$ and $\mathrm{W_2}$ of ribbon $\mathrm{R_2}$. These loops fan out toward the positive plage between $\mathrm{P_1}$ and $\mathrm{P_2}$.
\begin{figure*}[h]
\begin{interactive}{animation}{Figure6.mp4}
\epsscale{1.0}
\plotone{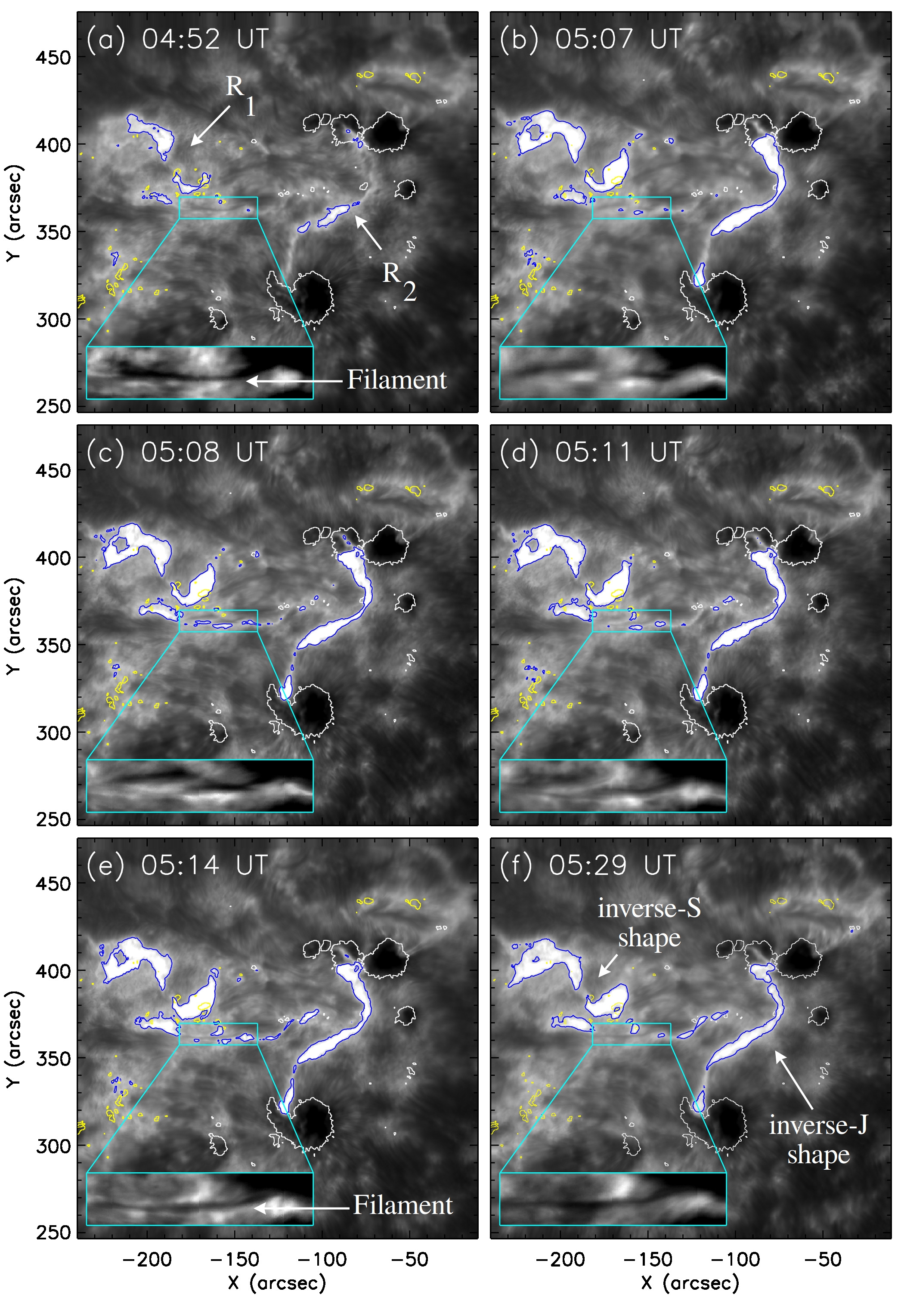}
\end{interactive}
\caption{Sequence of the MAST Ca II 8542\,\AA\,images of the second flare. The cyan box shows a zoomed-in view of the filament with enhanced image contrast. In selective panels, the filament, the inverse-S and inverse-J shapes of the flare ribbons $\mathrm{R_1}$ and $\mathrm{R_2}$ are pointed to by arrows. The white (yellow) contours trace $\mathrm{B}_z$=$\mathrm{+1000\,(-1000)\,G}$ at 03:00 UT, while the blue contours outline flare ribbons at intensity level 50,000 DN.\\ \textbf{Animation:} An animation showing the MAST images from 04:46 UT to 05:50 UT at a cadence of $\sim 32$ seconds is available in the HTML version of this article.
\label{mast}}
\end{figure*}
The sequential brightening of the footpoints of the loops in $\mathrm{L_1}$ and $\mathrm{L_2}$ along the ribbons at each end suggests that slipping reconnection heats the coronal flare loops in $\mathrm{L_1}$ and $\mathrm{L_2}$, and the ribbons $\mathrm{R_1}$ and $\mathrm{R_2}$ at their feet. We refer the reader to the animation for \autoref{131}, which clearly shows the sequential brightening of the footpoints of the flare loops in $\mathrm{L_1}$ and $\mathrm{L_2}$ along the ribbons of the twin flares.
Late in each flare, the southern extent of the $\mathrm{L_2}$ fan and its flare ribbon extends toward polarity $\mathrm{P_2}$. These southern bright loops have positive footpoints along the bottom part of the inverse-J shaped ribbon, while their negative ends are in $\mathrm{N_1}$. The full extent of the first flare becomes visible near the peak time of the flare, as in \autoref{131}(c). The second flare produces intense loops rooted in nearly the same ribbons, as in \autoref{131}(d). The loop footpoints tracing the bottom part of the inverse-J shaped ribbon seemingly move along the ribbon, indicating slipping reconnection. Here, it is important to remark that our use of the term ``slipping reconnection" is only used in the sense of a common feature of 3D reconnection, without ascribing to it the classification into ``slipping reconnection" and ``slip-running reconnection" that depends on whether the slippage speed is sub-Alfv\'enic or super-Alfv\'enic \citep{2006Au,2025ApJ...982L...9Z}, respectively. Nevertheless, we tracked the sequential brightenings of the footpoints of the loops in $\mathrm{L_1}$ and $\mathrm{L_2}$ along the segment $\mathrm{E_1}$ of $\mathrm{R_1}$ and the bottom part of the inverse-J shaped ribbon, respectively, from 03:39 UT to 03:53 UT at a cadence of one minute to estimate the speeds of the slipping motions. Using 10 repeated measurements at each time instant, the average speeds of the slipping motions during that time interval are found to be $21\pm3$ km/s and $46\pm6$ km/s, respectively. These estimates are of the same order as the speeds at which flaring loops have been reported to slip because of slipping reconnection in the papers of \citet{2014ApJ...784..144D}, \citet{Dudik_2016}, \citet{2016A&A...596A...1S}, and \citet{2020A&A...636L..11Z}. The tracked sequential brightenings are indicated in the animation of \autoref{131} by red boxes, whose horizontal and vertical extents represent the standard deviation in the measurements of their $x$ and $y$ coordinates, respectively. Importantly, both flares are morphologically nearly identical and hence, they are homologous---events that appear twice or multiple times in the same location with similar morphology \citep{1989SoPh..119..357M}. Furthermore, the precursors that made $\mathrm{T_1}$ and $\mathrm{T_2}$ each occur at the same site and are morphologically similar, and so are also homologous. Therefore, it is plausible that the triggers of the two homologous atypical flares were the two homologous precursors.

In addition to SDO/AIA, we used the line scan images of MAST in Ca II 8542\,\AA, sampled at 22 positions within $\pm\,0.7$\,\AA\, of the line center. The field of view is $3.8\arcmin\times3.8\arcmin$, with a temporal cadence of nearly 32 seconds over a duration of about 1\,hr\,15\,min, starting from 04:46 UT. Therefore, the MAST images on 2022 April 22 covered the second flare only. The images are rebinned from their original $\mathrm{2k \times 2k}$ size to a $\mathrm{1k \times 1k}$ size to increase the signal-to-noise ratio, while preserving the diffraction limit of the telescope. \autoref{mast} shows a sequence of six MAST images taken in the line center of the Ca II line at 8542\,\AA. The ribbons $\mathrm{R_1}$ and $\mathrm{R_2}$ are pointed to in \autoref{mast}(a) along with the filament for which a zoomed-in view is shown in the rectangular cyan box with enhanced contrast. The MAST images show that the filament's manifestation as a dark, elongated feature is disrupted during the filament's flaring dynamics, followed by the reappearance of a single coherent filament at 05:14 UT, as shown in \autoref{mast}(b)-\autoref{mast}(f). To show these changes clearly, all the panels include a zoomed-in view of the filament, as in \autoref{mast}(a). Furthermore, these changes are accompanied by brightening $\mathrm{C_2}$, thus supporting the interpretation that the activation of the filament resulted in the $\mathrm{C_2}$ pair of flare ribbons. Notably, $\mathrm{C_1}$ and $\mathrm{C_2}$ do not persist throughout the flaring dynamics, indicating that the filament does not play any role in the development of flare ribbons $\mathrm{R_1}$ and $\mathrm{R_2}$---a characteristic of atypical flares.
To investigate further, we computed a 3D magnetic field of the two active regions using magnetic field extrapolation, as discussed in the following.

\section{Magnetic Field Extrapolation}\label{extrapolation}

We extrapolated the active-region magnetic field using a nonlinear force-free field (NLFFF) model, which is based on a weighted optimization method \citep{W2021}. The details of this model can be found in \citet{W2010} and the references therein; here we provide a brief overview for completeness. The NLFFF model solves the equations $(\nabla \times {\textbf{B} }) \times{\textbf{B}} = \textbf{0}$ and $\nabla\cdot\mathbf{B} = 0$ with $\mathbf{B} = \mathbf{B}_{\mathrm{obs}}$ on the bottom boundary, where $\mathbf{B}_{\mathrm{obs}}$ denotes a vector magnetogram. This magnetogram is processed to make it compatible with the assumption of a force-free magnetic field. The processing of the magnetogram averages out small-scale features, thus smoothening the magnetogram. In the model formulation, the processed magnetogram is again labeled as $\mathbf{B}_{\mathrm{obs}}$ for convenience. Subsequently, the vertical component of $\mathbf{B}_{\mathrm{obs}}$ is used to compute a potential magnetic field for specifying the lateral and top boundaries of the computational box. The bottom boundary is replaced by $\mathbf{B}_{\mathrm{obs}}$. To minimize the effect of the side and top boundaries on the extrapolated field, they are replaced by boundary layers of finite width. Then, a functional given as
\begin{eqnarray}
L = \displaystyle\int_{V} w_{f}\frac{|(\nabla \times \mathbf{B})\times \mathbf{B}|^{2}}{B^{2}}\mathrm{d}^{3}x + \displaystyle\int_{V} w_{d}|\nabla \cdot \mathbf{B}|^{2}\mathrm{d}^{3}x +
\nu\displaystyle\int_{S}(\mathbf{B} - \mathbf{B}_{\rm obs})\ast\textbf{W}(x,y)\ast(\mathbf{B} - \mathbf{B}_{\rm obs})\mathrm{d}^{2}S
\label{1}
\end{eqnarray}
is minimized iteratively, where $w_f$ and $w_d$ are weight functions for the boundary layers \citep{W2010}, and ``$\ast$" denotes matrix multiplication . The surface integral term is evaluated over the bottom boundary only. In this term, $\mathbf{W}(x,y)$ is a diagonal matrix whose elements are inversely proportional to the measurement error in the magnetogram, and $\nu$ is a Lagrange multiplier that determines the extent to which the bottom boundary can relax \citep{Wiegelmann2012}.

The extrapolation is carried out at 03:00 UT using a vector magnetogram from the hmi.sharp\_cea\_720s data series. The magnetogram is rebinned to half of its original size, which increases the pixel size from 0.5\arcsec /pixel to 1\arcsec /pixel. The size of the computation box is $480\times320\times320$, corresponding to $348\times 232\times 232\,\mathrm{Mm^3}$ in physical units. The extrapolation uses a boundary layer of 32 voxels, resulting in an effective box size of $416\times 256\times 288$ in voxel units. The extrapolation in this study uses $w_f=w_d=1$ and $\nu=0.001$. To check how close the extrapolated magnetic field is to being force-free, we considered the current-weighted average angle ($\theta$) between current density ($\mathbf{J}$) and magnetic field ($\mathbf{B}$), defined by \citet{Wheatland2000} as
\begin{equation}
\theta = \mathrm{sin}^{-1}\sigma_{j},\,\,\,\,\sigma_{j}= \frac{\sum_{i}|\mathbf{J}|_{i}\sigma_{i}}{\sum_{i}|\mathbf{J}|_{i}},\,\,\,\,\sigma_{i} = \frac{|\mathbf{J}\times\mathbf{B}|_{i}}{|\mathbf{J}|_{i} \times |\mathbf{B}|_{i}}~, \label{angle-wheat}
\end{equation}
where $i$ runs over all the voxels in the computational box. The log files from the extrapolation reveal that $\theta=7.41^{\circ}$, which is acceptable, where we consider $\theta<10^{\circ}$ to be adequate for a force-free extrapolation. The log files also reveal that the extrapolated field's free magnetic energy is 11.4\% of the energy of the potential field.

For a morphological investigation of the extrapolated field, we calculated the squashing degree\footnote{The used numerical code is available at \href{http://staff.ustc.edu.cn/~rliu/qfactor.html}{http://staff.ustc.edu.cn/$\sim$rliu/qfactor.html}} (\textit{Q}), which quantifies gradients in mapping of magnetic field lines, thus allowing an identification of QSLs and HFTs \citep{2002TH}. The squashing degree is defined using the Jacobian matrix $D_{12}$ of the mapping $\prod_{12}$: $\vec{R}_{1}(x_{1},y_{1}) \mapsto \vec{R}_{2}(x_{2},y_{2})$ as
\begin{equation}
Q \equiv \frac{a^{2}+b^{2}+c^{2}+d^{2}}{|B_{n,1}(x_{1},y_{1})/B_{n,2}(x_{2},y_{2})|},\hspace{0.2em}D_{12}=\left[\frac{\partial\vec{R}_{1}}{\partial\vec{R}_{2}}\right] = \left(
\begin{array}{cc}
\partial x_2/\partial x_1 & \partial x_2/\partial y_1 \\
\partial y_2/\partial x_1 & \partial y_2/\partial y_1
\end{array}
\right)
=
\left(
\begin{array}{cc}
a & b \\
c & d
\end{array}
\right),
\end{equation}
where $\vec{R}_{1}(x_{1},y_{1})$, $\vec{R}_{2}(x_{2},y_{2})$ are the footpoints of a field line and $B_{n,1}(x_{1},y_{1})$, $B_{n,2}(x_{2},y_{2})$ are the vertical component of the magnetic field at the respective footpoints. QSLs are characterized by strong gradients in the mapping, having large \textit{Q} values. Furthermore, QSLs can be identified by finding those regions in which field lines that are close to each other at one location separate widely when going away from that location \citep{1997SoPh..175..123D}. In an HFT, two QSLs intersect \citep{2003ApJ...582.1172T}, resulting in an X-type cross section in the middle. 

Importantly, the cospatiality of features observed during a flare with magnetic structures in the extrapolated field is indicative of a relationship between the two, thus allowing a topological analysis of the flaring magnetic field. However, note that a comparison of the extrapolated fields at time $t_1$ with the flare brightenings at any other time $t_2$ requires the assumption that during the time window $t_2 - t_1$, the evolution of the magnetic flux and flare dynamics have not significantly impacted the shape and location of the reconnection sites or any other magnetic structures. This assumption is expected to hold acceptably well for the nearly two-hour duration of the flares in this study, at least for the field lines connecting strong magnetic polarities. 
In accordance with this notion, we used the extrapolated field to study the magnetic structure of the features observed during the two flares. 

\begin{figure*}[h]
\epsscale{1.0}
\plotone{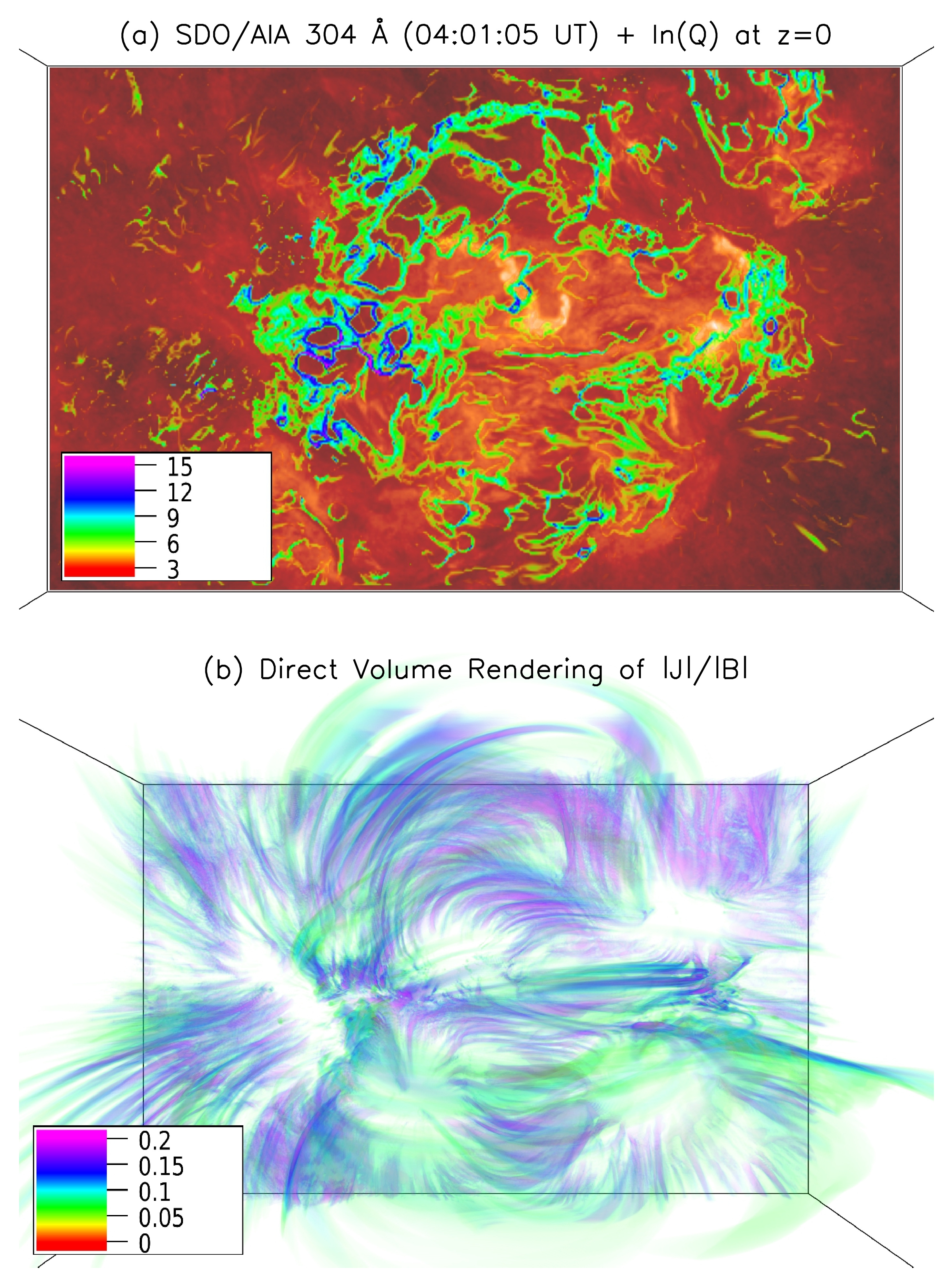}
\caption{Nonpotentiality of the NLFF model magnetic field. (a) Distribution of ln(\textit{Q}) in the range [3,15] on the bottom boundary of the computational domain, overlaid on the SDO/AIA 304\,\AA\,image at 04:01:05 UT. (b) Direct volume rendering (DVR) of $|\mathbf{J}|/|\mathbf{B}|$ in the range $[0,0.2]\,\mathrm{A\,m^{-2}\,T^{-1}}$, viewed along the downward direction along the \textit{z}-axis. The color bar in each panel scales the magnitude of ln(\textit{Q}) and $|\mathbf{J}|/|\mathbf{B}|$ values, respectively.
\label{nlf1}}
\end{figure*}

We used the VAPOR\footnote{Visualization and Analysis Platform for Ocean, Atmosphere, and Solar Researchers (VAPOR; \citealt{2019Atmos..10..488L})} software to visualize the structure of the extrapolated magnetic field, and quantities such as the squashing degree and the current density. In \autoref{nlf1}(a) of \autoref{nlf1}, the map of ln(\textit{Q}), ranging from ln(\textit{Q})=3 to ln(\textit{Q})=15, is shown on the bottom boundary (\textit{z}=0) of the computational box. The plot is overlaid with the SDO/AIA 304\,\AA\, image at 04:01:05 UT, whose opacity has been reduced to enhance the contrast of the ln(\textit{Q}) map. The map reveals that $\mathrm{ln}(Q)\gtrsim 5$ at many sites along the flare ribbons $\mathrm{R_1}$ and $\mathrm{R_2}$, but neither of the two ribbons is strongly cospatial with such values of ln(\textit{Q}). To examine whether this difference is due to the time difference between the extrapolated field (03:00 UT) and the SDO/AIA 304\,\AA\, image (04:01:05 UT), we performed an auxiliary NLFFF extrapolation at 04:00 UT, but it produced similar cospatiality as obtained at 03:00 UT, which suggests a limitation or shortcoming of the extrapolation model. Although the overlap between the flare ribbons and $\mathrm{ln}(Q)\gtrsim 5$ values is weak, it nevertheless indicates that the ribbons of the two atypical flares are made by reconnection at QSLs. Importantly, the ln(\textit{Q}) map gives no obvious indication of where the observed ribbons of the atypical flares will happen, as evident from \autoref{nlf1}(a), where $\mathrm{ln}(Q)\gtrsim 5$ in many places, both near to and away from the flare ribbons.

\autoref{nlf1}(b) shows the Direct Volume Rendering (DVR) of $|\mathbf{J}|/|\mathbf{B}|$---a proxy to locate regions of steep gradients of the magnetic field that are susceptible to reconnection \citep{2021ApJ...914...71I,2023PhyS...98f5016B,2024SoPh..299...15A}. The DVR feature of VAPOR facilitates the inspection of three-dimensional scalar fields through color mapping of their values. Furthermore, an opacity control enables adjustment of transparency, which can be used to visualize the scalar field at different depths within the computational volume. The plot shows $|\mathbf{J}|/|\mathbf{B}|$, ranging up to $0.2\,\mathrm{A\,m^{-2}\,T^{-1}}$, viewed along the downward direction along the \textit{z}-axis. Furthermore, lower values of $|\mathbf{J}|/|\mathbf{B}|$ have been suppressed, and the opacity has been adjusted such that the volumes having higher $|\mathbf{J}|/|\mathbf{B}|$ are visible. The plot reveals highest values ($\sim$0.2) of $|\mathbf{J}|/|\mathbf{B}|$ at multiple locations across the field of view, suggesting that the magnetic structures at all these locations have the same degree of non-potentiality. Therefore, assuming that this level of $|\mathbf{J}|/|\mathbf{B}|$ is sufficient to initiate reconnection would imply flaring activity at many locations. However, this is not observed; hence, the extrapolated field does not distinguish the flaring magnetic field from the surrounding non-flaring magnetic field. In other words, the magnetic field that is ready to reconnect and produce the observed flares is not obvious in the extrapolated field. Nevertheless, examining the extrapolated field in tandem with the flare observations enables some topological analysis of the two atypical flares.

\begin{figure*}[h]
\epsscale{1.0}
\plotone{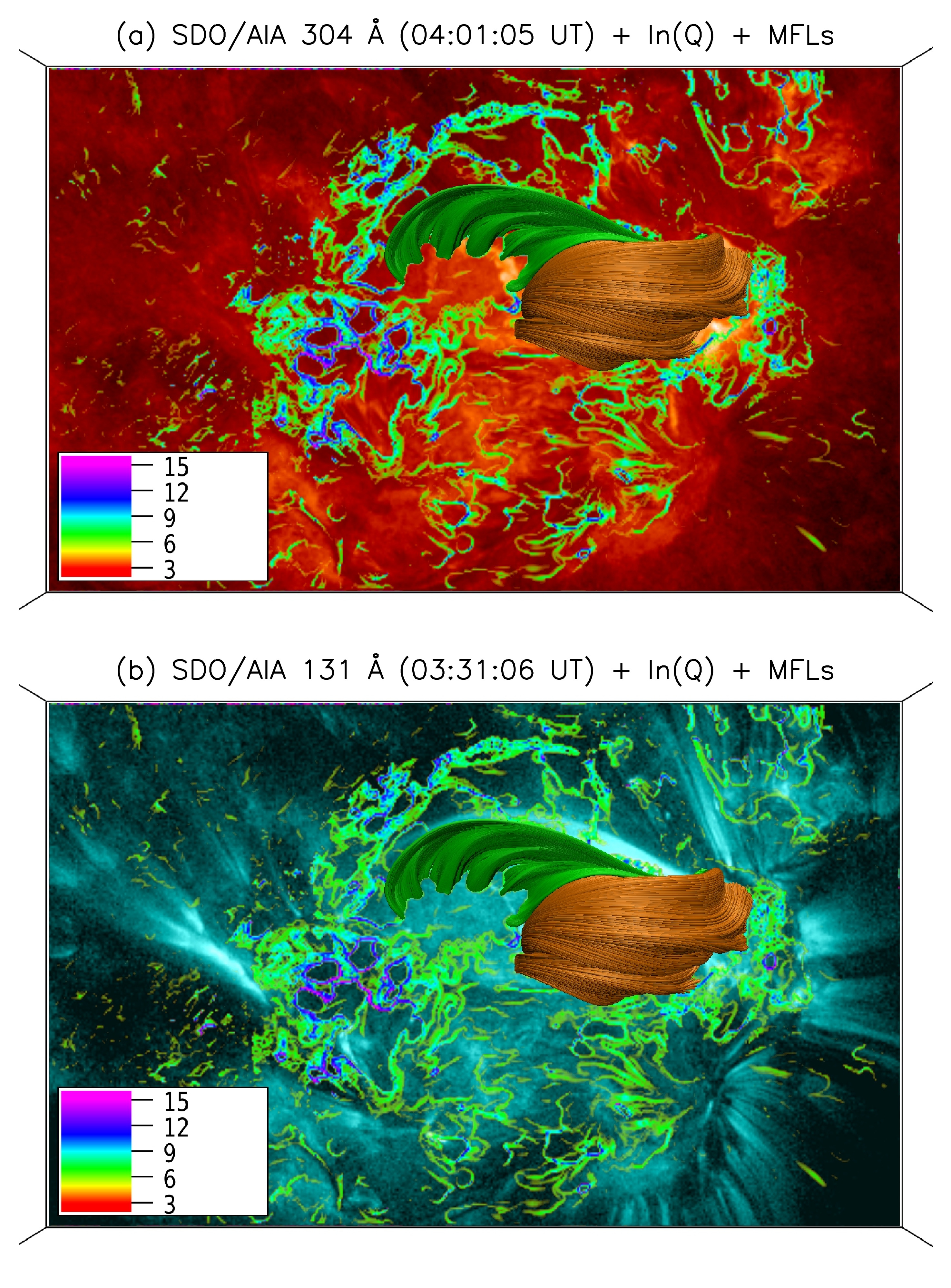}
\caption{Two quasi-separatrix layers with feet in and along the flare ribbons $\mathrm{R_1}$ and $\mathrm{R_2}$. The green and brown sets of magnetic field lines each display one of the QSLs. The field line plots are overplotted on the surface (photospheric) distribution of ln(\textit{Q}) in each panel, along with SDO/AIA 304\,\AA\, and 131\,\AA\,images at 04:01:05 UT and 03:31:06 UT in panels (a) and (b), respectively. The color bar in each panel scales the magnitude of ln(\textit{Q}).
\label{nlf2}}
\end{figure*}

\autoref{nlf2} depicts two sets of magnetic field lines (MFLs), in green and brown, overlaid on the SDO/AIA 304\,\AA\, and 131\,\AA\, images at 04:01:05 UT and 03:31:06 UT in panels (a) and (b), respectively. The green MFLs have closely spaced footpoints near $\mathrm{P_1}$ and fan out along the segment $\mathrm{E_1}$ of ribbon $\mathrm{R_1}$, thus making a QSL. The higher ln(\textit{Q}) values along $\mathrm{E_1}$, compared to the compact end near $\mathrm{P_1}$, further confirm our interpretation of this being a QSL. The brown MFLs are rooted in the segment $\mathrm{E_2}$ of ribbon $\mathrm{R_1}$ and near segments $\mathrm{W_1}$ and $\mathrm{W_2}$ of ribbon $\mathrm{R_2}$. These brown MFLs also fan out, but not as extremely as the green MFLs. However, since there is a clear spreading of these brown MFLs from the footpoints rooted in $\mathrm{E_2}$ to their opposite ends near ribbon $\mathrm{R_2}$, we infer that the brown MFLs also make a QSL whose compact end is rooted in segment $\mathrm{E_2}$ of ribbon $\mathrm{R_1}$. Furthermore, the ln(\textit{Q}) values along the footpoints rooted in $\mathrm{E_2}$ are smaller than those along the footpoints at the opposite end, implying that the brown MFLs indeed make a QSL. Also, the green and brown MFLs match the loop systems $\mathrm{L_1}$ and $\mathrm{L_2}$ in \autoref{131}(b), respectively. The green and brown MFLs do not show any field line connection that matches with the loop of the apparent X-shape (panel (a) of \autoref{131}) that is rooted in the plage northeast of $\mathrm{N_1}$ and in the page between $\mathrm{P_1}$ and $\mathrm{P_2}$. However, this does not mean there are no magnetic loops connecting the negative plage northeast of $\mathrm{N_1}$ to the positive plage between $\mathrm{P_1}$ and $\mathrm{P_2}$. The northeast to southwest loop of the X-shape in \autoref{131}(a) shows the presence of such magnetic loops. The model magnetic field lines rooted in and around $\mathrm{N_1}$ do not connect to the umbra of sunspot $\mathrm{P_2}$, presumably due to the asymmetry in the quadrupolar configuration. Even so, although the QSL made by the brown MFLs does not fan out into the bottom part of the inverse-J shaped ribbon, the observations imply that the actual QSL encompasses the entire ribbon. The green and brown model QSLs thus suggest that the field in the two flares consists of two QSLs that are rooted in ribbons $\mathrm{R_1}$ and $\mathrm{R_2}$, and are similar to the two model QSLs in \autoref{nlf2}.

\begin{figure*}[h]
\epsscale{1.0}
\plotone{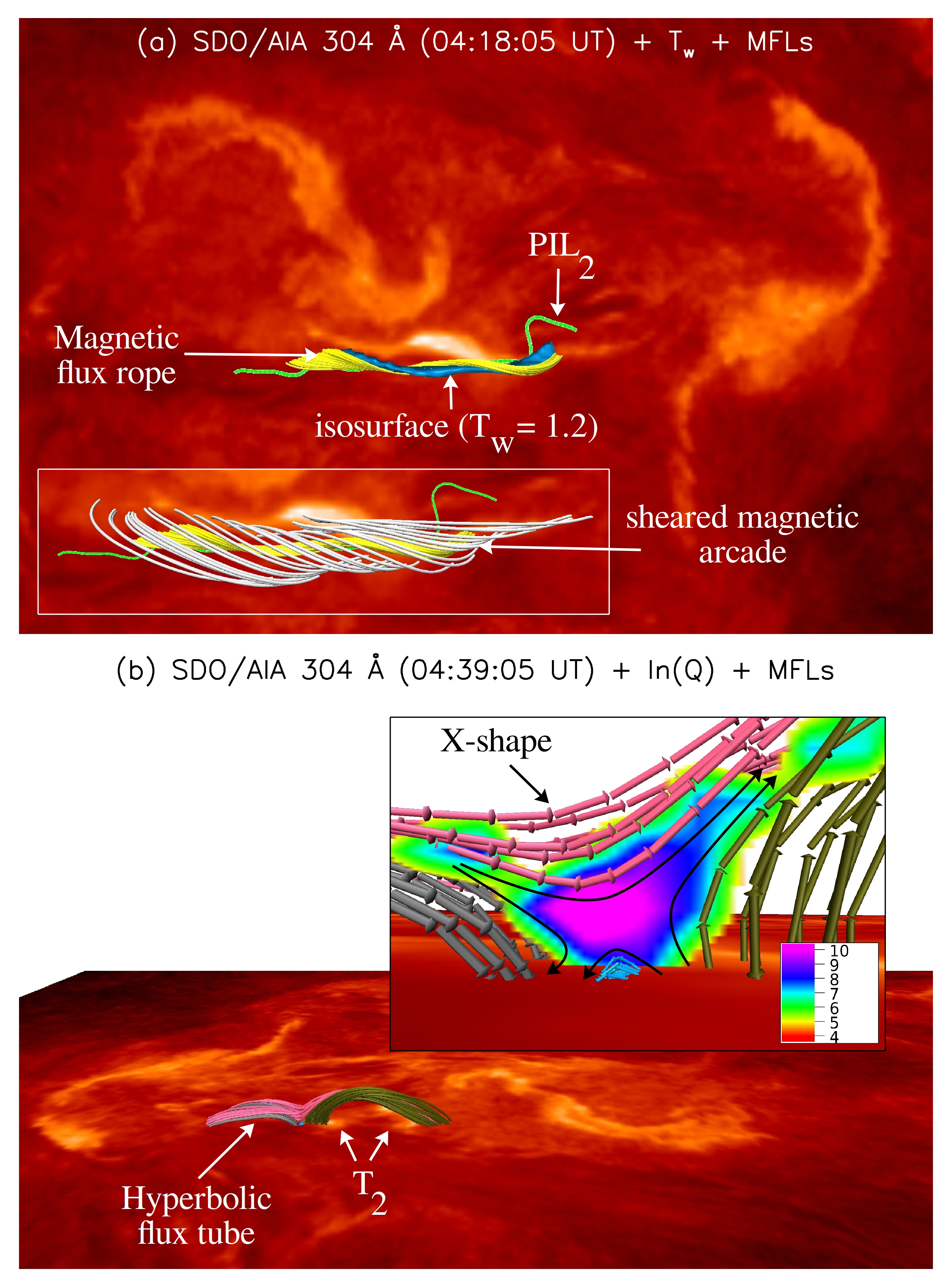}
\caption{Extrapolated magnetic field (a) at the site of $\mathrm{C_1}$ and (b) at the site of $\mathrm{T_2}$. (a) Magnetic flux rope along $\mathrm{PIL_2}$, shown by yellow MFLs and overlaid on the SDO/AIA 304\,\AA\,image at 04:18:05 UT. The isosurface in blue corresponds to $T_w = 1.2$, where $T_w$ is the twist number. The sub-panel shows a zoomed-in view of the MFR and the enveloping sheared arcade, shown by white MFLs (b) Hyperbolic flux tube at the site of ribbon pair $\mathrm{T_2}$, consisting of four connectivity domains, shown in blue, olive, pink, and grey MFLs. The plot is overlaid on the SDO/AIA 304\,\AA\,image at 04:39:05 UT. The sub-panel is a zoomed-in view of the X-shaped cross section of the HFT, showing field-line directions and an ln(\textit{Q}) map on a plane perpendicular to the bottom boundary. The curved black arrows further highlight the field-line direction and the X-shape. The color table scales the magnitude of ln(\textit{Q}), showing it has a maximum in the X-region of the HFT.
\label{nlf3}}
\end{figure*}

In addition to these QSLs, the extrapolated field has a magnetic flux rope that is cospatial with the filament, as depicted in panel (a) of \autoref{nlf3} by the yellow MFLs. The plot overlays the SDO/AIA 304\,\AA\,image at 04:18:05 UT, along with $\mathrm{PIL_2}$ and an isosurface for $T_w=1.2$, where $T_w$ is a measure of twist in the magnetic field lines and is given by $T_w=\int_{L}(\mu_{0}J_{\parallel})/(4\pi B)dl$, with $J_{\parallel}$ being the component of current density parallel to the magnetic field and $dl$ being an infintesimal length segment along a magnetic field line. The maximum value of $T_w$ in the MFR is nearly 1.5, which indicates that the MFR is moderately twisted. Furthermore, the MFR is inside a sheared magnetic arcade, as shown by the sub-panel in \autoref{nlf3}(a), where a zoomed-in view of the MFR is shown with enveloping magnetic arcade in white MFLs. Upon closer inspection of this sub-panel, it can be seen that some magnetic field lines in the arcade are rooted in the flare ribbons in the observed brightening $\mathrm{C_1}$. Furthermore, an auxiliary NLFFF extrapolation at 05:30 UT (the time up to which the filament remains identifiable) reveals the presence of a magnetic flux rope and an overlying sheared arcade at the location of the filament, suggesting that the magnetic configurations of $\mathrm{C_1}$ and $\mathrm{C_2}$ are similar. This configuration suggests that reconnections in the filament and its enveloping sheared magnetic arcade is the most plausible explanation for the observed brightenings $\mathrm{C_1}$ and $\mathrm{C_2}$. However, since our primary focus is on the ribbons of the two atypical flares, namely $\mathrm{R_1}$ and $\mathrm{R_2}$, we do not examine the reconnection making $\mathrm{C_1}$ and $\mathrm{C_2}$ any further.

Lastly, we investigated the magnetic field configuration for ribbons $\mathrm{T_1}$ and $\mathrm{T_2}$, and found a hyperbolic flux tube cospatial with them. The HFT is shown in panel (b) of \autoref{nlf3}, and consists of four connectivity domains, shown in blue, olive, pink, and grey MFLs. The plot overlays the SDO/AIA 304\,\AA\,image at 04:39:05 UT and shows a zoomed-in view of the HFT, displaying the X-shaped cross section---a characteristic property of any HFT. The zoomed-in plot shows the direction of magnetic field lines along with a map of ln(\textit{Q}) in a plane perpendicular to the bottom boundary and passing through the X-shaped cross section of the HFT. Usually, an X-shaped cross section is revealed by the distribution of ln(\textit{Q}) values, but in our case, the central region (ln(\textit{Q})$\geq 10$) of the X-shape is very close to the bottom boundary and consequently, the X-shape in the ln(\textit{Q}) map is only partially visible. We have drawn curved black arrows in the sub-panel to further indicate the direction of MFLs and the X-shaped cross-section of the HFT.
As for $\mathrm{C_1}$ and $\mathrm{C_2}$, we do not examine the formation of $\mathrm{T_1}$ and $\mathrm{T_2}$ any further. From their times however, we infer that $\mathrm{T_1}$ and $\mathrm{T_2}$ are homologous precursors that possibly triggered the twin atypical flares. For the two atypical flares and their flare ribbons, we propose a scenario for the magnetic reconnection that makes them, which is discussed in the following section.

\section{Proposed Flare Scenario}\label{pfs}

From our analysis of the flare images and the extrapolated magnetic field, we surmise that there are essentially two QSLs rooted in the flare ribbons of the twin atypical flares. We propose that slipping reconnection in these two QSLs heats the flare ribbons $\mathrm{R_1}$ and $\mathrm{R_2}$, and the magnetic loops rooted in them.
First, we note that in slipping reconnection, the connectivity of field lines changes in a continuous and sequential manner. In other words, unlike the ``cut" and ``paste" reconnection in a strictly two-dimensional current sheet, slipping reconnection is characterized by a continuous sequential change in the connectivity of magnetic field lines. This continuous change in the connectivity of field lines implies that reconnection proceeds sequentially in space and time, involving successive reconnection of neighboring field lines. That is, during slipping reconnection, a series of magnetic reconnections occur, involving numerous pairs of crossed magnetic field lines. The reconnection of each pair changes the field-line footpoint by only a small amount, because the field lines cross at small angles. Notably, the sequential pairs of field lines include not only those present at the onset of reconnection but also those that form during the process, producing reconnected and yet-to-reconnect field lines.
The field lines cross each other in the 3D space and undergo component reconnection, which refers to reconnection between magnetic field lines that are not exactly anti-parallel. ``Slipping reconnection," which produces what seems to be movement of the footpoints of reconnected magnetic field lines along the flare ribbons, can therefore be envisaged as a manifestation of many sequential component reconnections in a localized (in our case, shell-like) non-ideal region within each of our two QSLs.

\begin{figure*}[h]
\epsscale{1.0}
\plotone{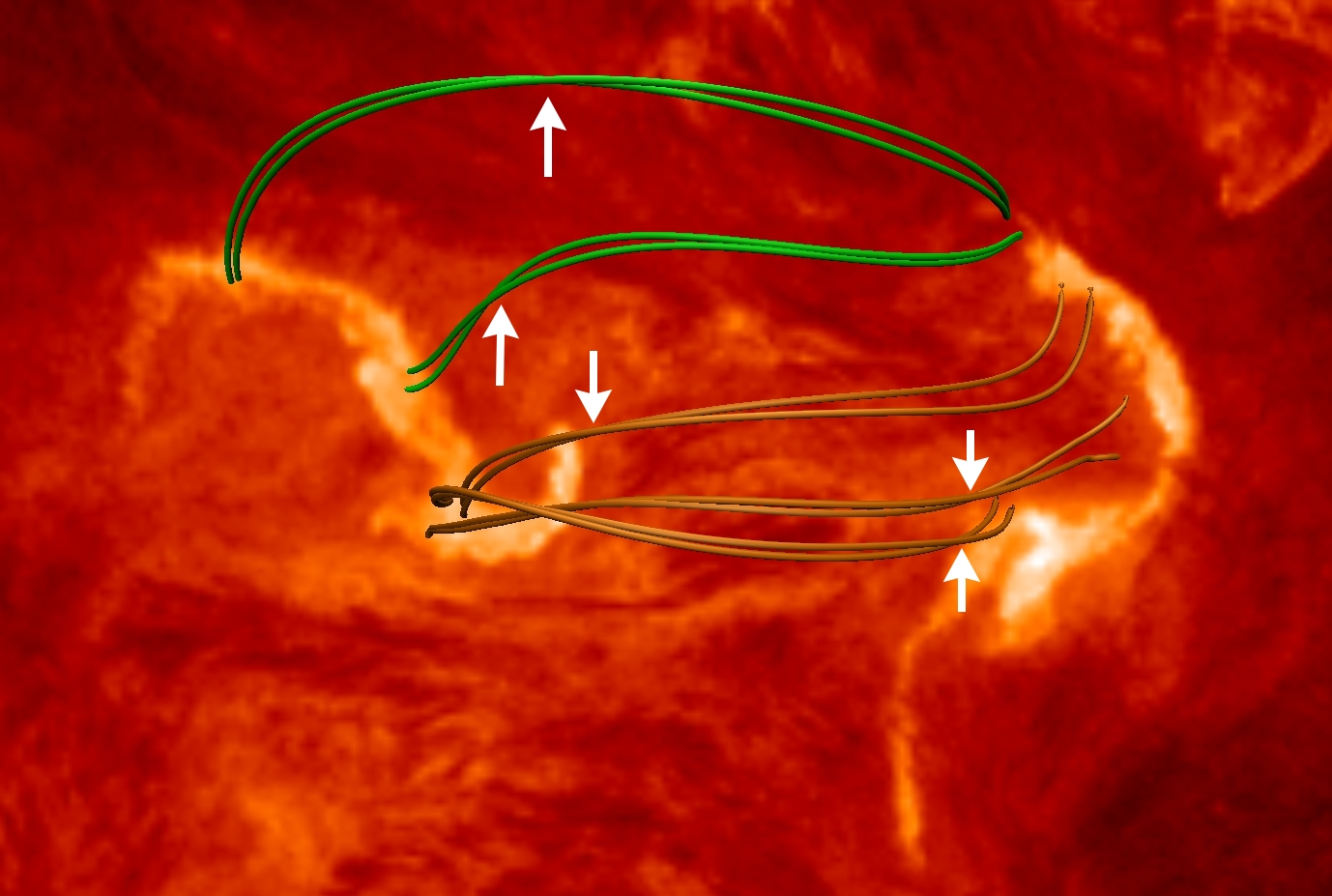}
\caption{Selected pairs of slightly crossed magnetic field lines within the QSLs shown in \autoref{nlf2}. The plot is on the SDO/AIA 304\AA\,\,image at 04:01:05 UT, with white arrows pointing to apparent crossing points between the field lines.
\label{last}}
\end{figure*}

In accordance with this concept, we attempted to find within the two QSLs pairs of closely spaced extrapolated magnetic field lines, exhibiting a crossing point and having feet in the flare ribbons $\mathrm{R_1}$ and $\mathrm{R_2}$. For an actual crossing point in the 3D space, its coordinates must be nearly preserved when the field lines are projected onto a 2D plane from orthogonal directions. In our extrapolated QSLs, we found pairs of magnetic field lines that seem to cross each other when viewed downward along the \textit{z}-axis. Some of the identified pairs are depicted in \autoref{last} as examples, where such crossing points are pointed to by the solid white arrows. However, all the crossing points and their coordinates were not found to be preserved when viewed from the positive \textit{y}-axis, meaning that either the crossing points disappeared or the change in their coordinates was significant. This suggests that the field lines corresponding to such crossing points are not close enough to reconnect. For such crossing points, our attempt to find a pair of field lines having a close crossing yielded lines that are inclined at extremely small angles, making it difficult to show a crossing point clearly. We think that this reflects the fact that the extrapolated nonlinear force-free field does not distinguish the flaring region from the surrounding non-flaring regions, which implies that the required degree of non-potentiality or misalignment between field lines is not captured by the extrapolation.

\begin{figure*}[h]
\epsscale{1.0}
\plotone{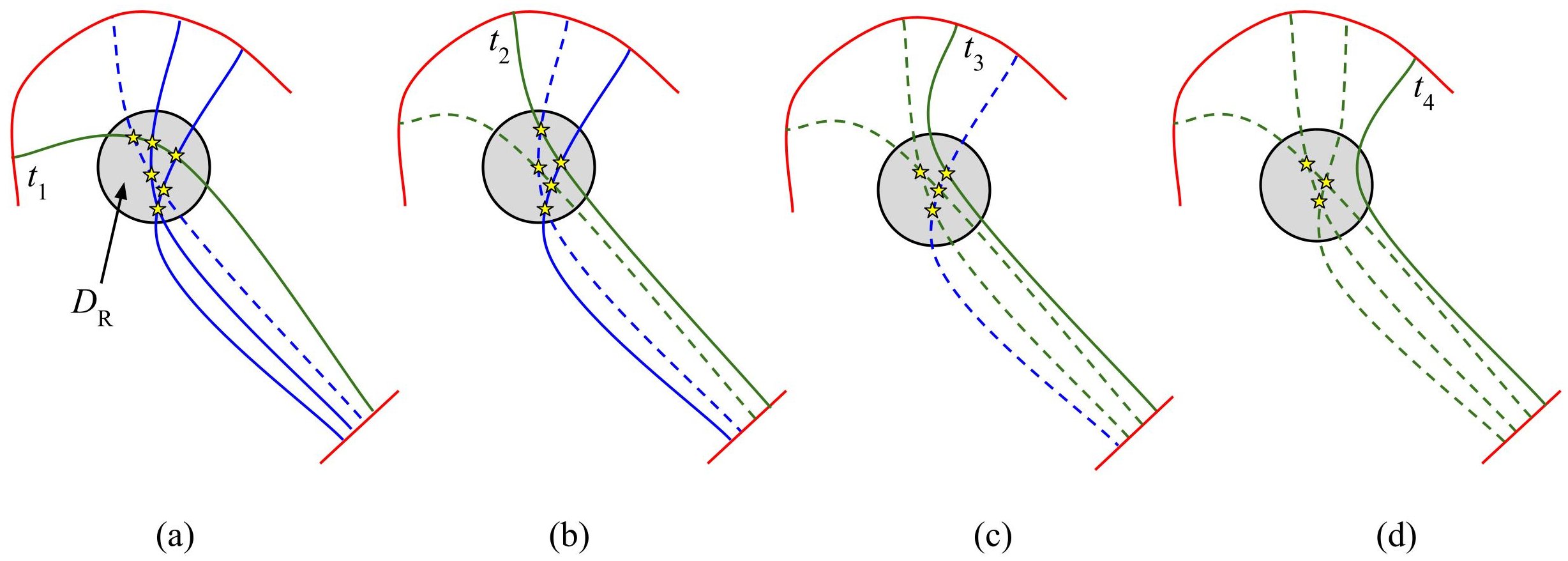}
\caption{Schematic of slipping reconnection by sequential component reconnections between misaligned magnetic field lines. In each panel, the solid green line reconnects with the dashed blue line, while the dashed green line represents the post-reconnection state of the dashed blue line. The solid blue and solid red lines denote the yet-to-reconnect field lines and the flare ribbons, respectively. The grey circle is the localized non-ideal (diffusive) region (labeled $D_{\mathrm{R}}$), and the yellow stars mark the apparent crossing points between the magnetic field lines. Panels (a)–(d) show the configuration at four successive time instants, from $t_1$ to $t_4$. 
\label{test1}}
\end{figure*}

Nevertheless, we think our interpretation that slipping reconnection in both QSLs (see \autoref{nlf2}) makes the two atypical flares remains plausible, where the slipping reconnection is understood to be a result of sequential component reconnections in adjacent pairs of magnetic field lines. To further illustrate our idea, a schematic is shown in \autoref{test1}. In each panel, the solid green line reconnects with the dashed blue line. The dashed green line depicts the dashed blue line with the same stronger flux footpoint after reconnection. The solid blue and solid red lines represent the yet-to-reconnect field lines and flare ribbons, respectively. The grey circle is the localized non-ideal region (labeled $D_{\mathrm{R}}$) and the yellow stars mark the crossing points between the magnetic field lines. In the schematic, we show sequential reconnections of the solid green line only. \autoref{test1}(a) to \autoref{test1}(d) show the configuration at four different times, from $t_1$ to $t_4$. The solid green line is shown to reconnect three times, and in each reconnection, its footpoint on the end of widely separated field-line footpoints takes a step along the flare ribbon. These steps produce the apparent slipping motion of slipping reconnection.

Finally, we summarize our overall scenario for the production of the twin pair of atypical flares and the pairs of flare ribbons observed in addition to the flare ribbons made by the atypical flares. We speculate that initially, the low-lying HFT near $\mathrm{P_{1}}$ undergoes sudden reconnection, heating $\mathrm{T_1}$. The reconnection in this HFT might have been triggered by flux emergence/cancellation or stressing of the HFT by photospheric convection. As a result of this event, the equilibrium of the neighboring magnetic structures is disturbed, triggering slipping reconnection in both QSLs. The slipping reconnection in the QSL shown by the green MFLs makes the segment $\mathrm{E_1}$ of ribbon $\mathrm{R_1}$, the inverse-S shaped ribbon. The segment $\mathrm{E_2}$ of ribbon $\mathrm{R_1}$, and ribbon $\mathrm{R_2}$ (inverse-J shaped ribbon) are made by slipping reconnection in the QSL shown by brown MFLs. During the decay phase of the first flare, the ribbons in brightening $\mathrm{C_1}$ are made by reconnections in the filament and its enveloping sheared magnetic arcade. Following this, the above-described scenario of reconnections is repeated with the precursor event $\mathrm{T_2}$ triggering the GOES M1.1 flare.

\section{Summary and Discussion}\label{sad}

In this study, we have investigated a twin pair of atypical flares (GOES C7.7 and GOES M1.1) that occurred on 2022 April 22 in a quadrupolar magnetic configuration (having an inverse-S shaped PIL) and constituted by AR 12993 ($\mathrm{P_1},\mathrm{N_1}$) and AR 12994 ($\mathrm{P_2},\mathrm{N_2}$), the terms inside the parantheses denoting the magnetic polarities. The two ARs together have only three clusters of full-fledged sunspots, making the configuration asymmetric, which manifests in little or no magnetic connectivity between $\mathrm{P_2}$ and $\mathrm{N_1}$. Furthermore, the extended plages have fragmented flux. The prior-day evolution of the photospheric magnetic flux shows that the flux steadily decreases, suggesting that flux cancellation dominates over flux emergence and transport of flux to outside the field of view. However, the role of this flux cancellation in the triggering of the two flares, if any, remains unclear. The analysis of images from the 304\,\AA\,\,and 131\,\AA\,\,wavelength channels of SDO/AIA, and the Ca II 8542\,\AA\,\,line of the ground-based Multi-Application Solar Telescope reveal several key features of the flaring dynamics. We identify five pairs of flare ribbons. The ribbons of the two atypical flares, labeled as $\mathrm{R_1}$ and $\mathrm{R_2}$, exhibit an inverse-S shape for $\mathrm{R_1}$ and an inverse-J shape for $\mathrm{R_2}$. The ribbon $\mathrm{R_1}$ is in the plage in and near $\mathrm{N_1}$ while $\mathrm{R_2}$ is in the plage between $\mathrm{P_1}$ and $\mathrm{P_2}$. Furthermore, both of the ribbons appear segmented into two parts, namely $\mathrm{E_1}$ and $\mathrm{E_2}$ in $\mathrm{R_1}$, and $\mathrm{W_1}$ and $\mathrm{W_2}$ in $\mathrm{R_2}$. This segmentation is perhaps due to the fragmentation of the magnetic flux.
The ribbons $\mathrm{R_1}$ and $\mathrm{R_2}$ do not spread apart, which signifies that the twin flares are atypical. In addition to the flare ribbons $\mathrm{R_1}$ and $\mathrm{R_2}$, we find four pairs of flare ribbons that each last only a few minutes, and each of which is from a different energy-release event. One of these four pairs, namely $\mathrm{T_1}$, happens just before the first atypical flare. At the location of $\mathrm{T_1}$, another pair of ribbons, namely $\mathrm{T_2}$, occurs at or soon before the start of the second atypical flare. The timings and locations of $\mathrm{T_1}$ and $\mathrm{T_2}$ relative to the atypical flares suggest that $\mathrm{T_1}$ and $\mathrm{T_2}$ each respectively trigger the first and second atypical flares. This, however, is not definitive evidence that these events triggered the onset of the two atypical flares, and our work does not rule out the possibility that $\mathrm{T_1}$ and $\mathrm{T_2}$ are random small-scale brightenings that by chance both occurred prior to the respective atypical flares.

The other two pairs of ribbons, namely $\mathrm{C_1}$ and $\mathrm{C_2}$, appear during the gradual and impulsive phases of the first and second flares, respectively, and bracket the filament along $\mathrm{PIL_2}$. The filament gets activated during each of the atypical flares, which activations accompany the brightenings $\mathrm{C_1}$ and $\mathrm{C_2}$.

The observations show two distinct loop fans, marked $\mathrm{L_1}$ and $\mathrm{L_2}$, rooted in the flare ribbons $\mathrm{R_1}$ and $\mathrm{R_2}$. The positive footpoints of the loops in fan $\mathrm{L_1}$ are compactly clustered near $\mathrm{P_1}$, while their negative footpoints fan out into the plage northeast of $\mathrm{N_1}$. Similarly, the loops in fan $\mathrm{L_2}$ are rooted in segment $\mathrm{E_2}$ of the inverse-S shaped ribbon and in segments $\mathrm{W_1}$ and $\mathrm{W_2}$ of the inverse-J shaped ribbon. The sequential brightening of the footpoints of these loops along the flare ribbons indicates that slipping reconnection heats ribbons $\mathrm{R_1}$ and $\mathrm{R_2}$. 

To study the magnetic field tracing and rooted in the observed features, we extrapolated the coronal magnetic field using an NLFFF model, from a photospheric a vector magnetogram at 03:00 UT as the bottom boundary. The extrapolation shows an HFT at the site of $\mathrm{T_1}$ and $\mathrm{T_2}$, a magnetic flux rope with an enveloping sheared arcade along $\mathrm{PIL_2}$, and two QSLs rooted in and near the flare ribbons $\mathrm{R_1}$ and $\mathrm{R_2}$. The observations and the extrapolated field together suggest that reconnection at the HFT makes $\mathrm{T_1}$ and $\mathrm{T_2}$, and that those eruptions each triggered one of the twin flares. The observations and model field also suggest that reconnection in the filament and the overlying arcade makes $\mathrm{C_1}$ and $\mathrm{C_2}$, and that the occurrence of slipping reconnection simultaneously in the two QSLs makes the twin pair of atypical flares.
We propose that slipping reconnection in the QSLs is many component reconnections of many pairs of magnetic field lines crossed at small angles. Overall, we find the twin pair of flares to be atypical, and we corroborate previous findings that indicate slipping reconnection makes atypical flares. In addition, some other aspects of the atypical flares investigated in this study are as follows. 

A hyperbolic flux tube can form in a quadrupolar configuration consisting of two bipolar sunspot groups \citep{2002TH,2003ApJ...582.1172T} separated by an S-shaped inversion line. However, our two active regions result in an asymmetric quadrupolar geometry that does not contain a large-scale HFT. Furthermore, because the flare brightenings encompass the polarities $\mathrm{P_1}$, $\mathrm{N_1}$, and $\mathrm{P_2}$ only, reconnection at a single large-scale HFT cannot account for the absence of flaring emission in polarity $\mathrm{N_2}$. Also, each atypical flare spans brightenings accompanying the activation of a filament, which we take to be confined standard flares (the case of \citealp{2001ApJ...552..833M}, Figure 1, lower-left panel), where the filament rises only a very small amount before being confined, and therefore where the flare ribbons move an imperceptible amount. In this way, our observations here are a case in which flare brightenings made by standard confined flares (in the form of the filament activations) and non-standard confined flares (the two atypical flares) are observed simultaneously within one complex magnetic field region. However, since the emission is dominated by the atypical flares, and since the flaring events that made $\mathrm{C_1}$ and $\mathrm{C_2}$ occurred during the two atypical flares, the GOES X-ray profile does not show the emission from $\mathrm{C_1}$ and $\mathrm{C_2}$. Further, since the two atypical flares are morphologically nearly indentical and since each of them is plausibly triggered by a burst of reconnection at the low-lying HFT near $\mathrm{P_1}$, the twin flares are homologous, and plausibly have homologous triggers.

In closing, we put forth some ideas on atypical flares that we believe merit investigation in future studies. Previous studies of atypical flares have proposed a few ideas to explain their triggering and driving mechanisms. For example, while we find that flux cancellation dominates the temporal evolution of the magnetic flux in the quadrupolar configuration that hosted the atypical flares reported in the present study, \citet{2015A&A...574A..37D} identify an episode of flux emergence as the trigger and continued flux emergence as the driver of the atypical flare in their study. Furthermore, while we find that the precursor events $\mathrm{T_1}$ and $\mathrm{T_2}$ possibly triggered the observed atypical flares, studies of additional atypical flares should be carried out; if brightenings of the nature we find here occurring immediately prior to the onset of the atypical flares turn out to be common, then that will support that the events ($\mathrm{T_1}$ and $\mathrm{T_2}$) we observe here are true precursors. Furthermore, if such evidence is found, it will be of high interest to identify the magnetic connections between the precursor events and the atypical flares, to clarify the triggering mechanism. \citet{2019ApJ...871..165J} did not report anything regarding the triggering of the atypcial flare in their study, while \citet{2019ApJ...881..151L} argued that multiple slipping reconnections made the atypical flare investigated in their study. Building on the remark in \citet{2022ApJ...933..191D}, stating that the triggering mechanism and the development process of atypical flares are unknown, we put forward a possibility based on the study by \citet{1997SoPh..175..123D}. That study posits that the photospheric magnetic field is fragmented into a large number of thin flux tubes, implying the presence of a large number of very thin QSLs in the solar corona. They argue that the photospheric motions create thin current layers at these QSLs, ultimately leading to a breakdown of ideal MHD, and hence magnetic reconnection. Importantly, they propose that such a process is common to both coronal heating and flares, depending on the amount of fragmentation in a region. Since the photospheric field in plages is highly fragmented, and since atypical solar flares seem to occur mainly in regions of weak magnetic field such as plages (see \citealt{2015A&A...574A..37D}, \citealt{2019ApJ...871..165J}), we think that the process of energy release proposed by \citet{1997SoPh..175..123D} merits investigation in the context of atypical flares. 

Atypical flares are characterized by the presence of stable filaments. However, the role of filaments during the flaring activity is unclear. For example, \citet{2015A&A...574A..37D} did not report anything in this regard, while \citet{2019ApJ...871..165J} found the filament to initiate reconnection in the overlying loops. Furthermore, in \citet{2022ApJ...933..191D}, stability of the filament is argued to be a possible consequence of small twist or strong strapping force of the overlying field lines. We think that both of these possibilities are relevant to the case of ``Type-II" flares. In our study, we observe that the filament gets activated during both atypical flares. Moreover, the formation of flare ribbons in atypical flares is dictated by QSLs, not by an erupting filament/magnetic flux rope. In consideration of these results, we propose that the presence of a filament is irrelevant for most atypical flares. Further studies are required to further assess the role of filaments and to better constrain the definition of atypical flares.


For the two atypical solar flares that we investigated in this study, we suppose that the QSLs made by the green and brown MFLs (see \autoref{nlf2}) each has many thin QSLs inside it. We also suppose that each of the two crossed loops making the X-shape in \autoref{131}(a) is being heated by internal slipping reconnection, not by external reconnection of each loop with the other.

We also note that our scenario for the ``atypical flares" in this study is essentially similar to that proposed by \citet{2024ApJ...975...20M} for the ``stealth non-standard-model confined flares" investigated in their study. Specifically, our scenario of simultaneous occurrence of several component reconnections between magnetic field lines crossing each other at small angles within each QSL is similar to that of sudden reconnection between magnetic field strands crossing each other at small ($10^{\circ}-20^{\circ}$) angles in seemingly inert magnetic arches. This similarity suggests that both ``atypical flares" and ``stealth non-standard-model confined flares" are made by the same process, though their particular details may vary depending on the  complexity of the magnetic configuration, thus requiring further investigations.

Finally, our scenario for the energy release in the twin pair of atypical flares reported here gives no reason for the observed placement and ribbon form of the conjugate flare ribbons. Apparently, the slipping reconnection occurs in a warped shell of coronal magnetic field that is rooted in the conjugate flare ribbons. How and why was that particular shell prepared for the slipping reconnection that made the twin flares? How and why did that shell release free magnetic energy in two consecutive flares instead of in only a single flare? Those remain open questions.
\begin{acknowledgments}
We thank the SDO team for their open data policy, which allowed us to freely use the data products from AIA and HMI instruments. We thank the MAST operations team for helping in the observations. We acknowledge the National Center for Atmospheric Research’s Computational and Information Systems Lab for their open-source visualization tool VAPOR, which allowed us to construct the magnetic field line, slice rendering, and DVR plots. S.A., A.C.S., and R.L.M. received funding from the Heliophysics Division of NASA’s Science Mission Directorate through the Heliophysics Supporting Research (HSR) Program. S.A. and Q.H. acknowledge support from NASA grants 80NSSC21K0003 and 80NSSC21K1671, and NSF grants AST-2204385 and AGS-1954503. S.A. thanks David Pontin for a discussion on the work. We would also like to thank the anonymous referee for insightful suggestions
and comments, which improved the overall quality of the paper.

\end{acknowledgments}


\bibliography{mybib}{}
\bibliographystyle{aasjournalv7}



\end{document}